\documentclass[aps, prd, showpacs, showkeys ]{revtex4}

\usepackage[small]{caption}
\usepackage{amsmath,amsfonts,amscd,amssymb,epsf, epsfig}\usepackage{graphicx}

\newcommand{\pa}{\partial}

\begin{document}

 \title{Casimir interaction between a sphere and a cylinder}

\author{L. P. Teo}
 \email{LeePeng.Teo@nottingham.edu.my}
\affiliation{Department of Applied Mathematics, Faculty of Engineering, University of Nottingham Malaysia Campus, Jalan Broga, 43500, Semenyih, Selangor Darul Ehsan, Malaysia.}
\begin{abstract}
We study the Casimir interaction between a sphere and a cylinder both subjected to Dirichlet, Neumann or perfectly conducting boundary conditions. Generalizing the operator approach developed by Wittman [IEEE Trans. Antennas Propag. \textbf{36}, 1078 (1988)], we compute the scalar and vector translation matrices between a sphere and a cylinder, and thus write down  explicitly the exact  TGTG formula for the Casimir interaction energy. In the scalar case, the formula shows manifestly that the Casimir interaction force is attractive at all separations. Large separation leading term of the Casimir interaction energy is computed directly from the exact formula. It is of order $\sim \hbar c R_1/[L^2\ln(L/R_2)]$, $\sim \hbar c R_1^3R_2^2/L^6$ and $\sim \hbar c R_1^3/[L^4\ln(L/R_2)]$ respectively for Dirichlet, Neumann and perfectly conducting boundary conditions, where $R_1$ and $R_2$ are respectively the radii of the sphere and the cylinder, and $L$ is the distance between their centers.
\end{abstract}
 \maketitle
 \section{Introduction}
 Casimir effect \cite{1} has attracted a lot of attention due to its wide applications in different areas of physics and its potential impact to nanotechnology. For a good review about the subject, one can read for example the book \cite{2}.

Naively,  the Casimir energy is a sum of the ground state energies of all the eigenmodes of the quantum field. However, the computation of the Casimir energy is not a simple task since the naive summation is divergent  and regularization is required.  Before the turn of the century, the exact computations were limited to some simple configurations with a particularly large amount of works being devoted to the configuration of two parallel plates.

Since the work of Lamoreaux \cite{41} in 1997, Casimir force have been measured with high precision in various configurations \cite{42}. This has   stimulated theorists to  study the  Casimir force between any two objects. At the beginning, several approximation schemes were developed for this purpose, such as the semiclassical approach \cite{3,4, 5}, the optical path approximation \cite{6, 7, 8} and the multiple reflection approximation \cite{9}.

Since the year 2006, exact computation of the Casimir energy between two objects has become  possible.
In \cite{10,11,12,13,14}, Gies et al. derived a worldline representation of the Casimir interaction between two objects imposed with Dirichlet boundary conditions.   In \cite{15}, Bulgac et al. computed the Casimir interaction energy between Dirichlet spheres or between a Dirichlet sphere and a Dirichlet plate using the multiple scattering approach, whose application in Casimir effect can be dated back to the work of  Balian and   Duplantier \cite{16,17}.
 Later   several other researchers  computed the Casimir interaction in various  configurations such as sphere-sphere, sphere-plane, cylinder-plane and cylinder-cylinder \cite{18,19,20,21,22,23,24,25,26,27,28,32,33,34,35} using some form of multiple scattering approach. A general scheme for computing the Casimir interaction between any two objects was developed by Emig et al. in \cite{21} for scalar fields and in \cite{24} for electromagnetic fields.

A totally different approach was used by Dalvit et al.  \cite{29,30,31}. They used mode summation approach  to compute the Casimir interaction energy of two eccentric cylinders with Dirichlet, Neumann or perfectly conducting boundary conditions.  Recently, we developed a general scheme for computing the Casimir interaction energy between two objects from the perspective of mode summation approach \cite{37}, which is fundamentally equivalent to the scheme developed by Emig et al. from the perspective of multiple scattering theory.

The basic ingredients in the exact representation of the Casimir interaction energy between two objects are the T-matrices of each of the objects and the translation matrices between the objects. The T-matrix of an object only depends on the coordinate system chosen and the boundary conditions imposed on that object, and it has been calculated for a plane, a sphere and a cylinder under various boundary conditions. The more difficult part in obtaining the exact representation of the Casimir interaction energy is in the computation of the translation matrices. For cylinder-cylinder and sphere-sphere configurations, the translation matrices are well known. For cylinder-plane and sphere-plane configurations, they have been obtained in \cite{24,37}. To the best of our knowledge, the translation matrices for the sphere-cylinder configuration have not been worked out before, and hence the Casimir interaction between a sphere and a cylinder has never been studied. The goal of the current work is to fill in this gap.

In Section \ref{sec2}, we generalize the operator approach in \cite{36}  to compute the translation matrices between a sphere and a cylinder for a scalar field, and hence write down the TGTG (exact) formula for the Casimir interaction energy between a sphere and a cylinder which are both imposed with Dirichlet boundary conditions or Neumann boundary conditions. In Section \ref{sec3}, we do the same for electromagnetic Casimir interaction between a perfectly conducting sphere and a perfectly conducting cylinder. In Section \ref{sec4}, we use the exact TGTG formulas to derive the leading term of the Casimir interaction energy when the separation between the sphere and the cylinder is large. In the Appendix \ref{sec5}, we  use the proximity force approximation \cite{39,40} and the derivative expansion proposed by  \cite{43} to compute  the small separation leading order term  and next to leading order term of the Casimir interaction.  In principle, these can also be computed from the exact formula but it is too complicated to be included here. This problem will be addressed in the future.

   \section{Exact scalar Casimir interaction energy}\label{sec2}
   \begin{figure}[h]
\epsfxsize=0.6\linewidth \epsffile{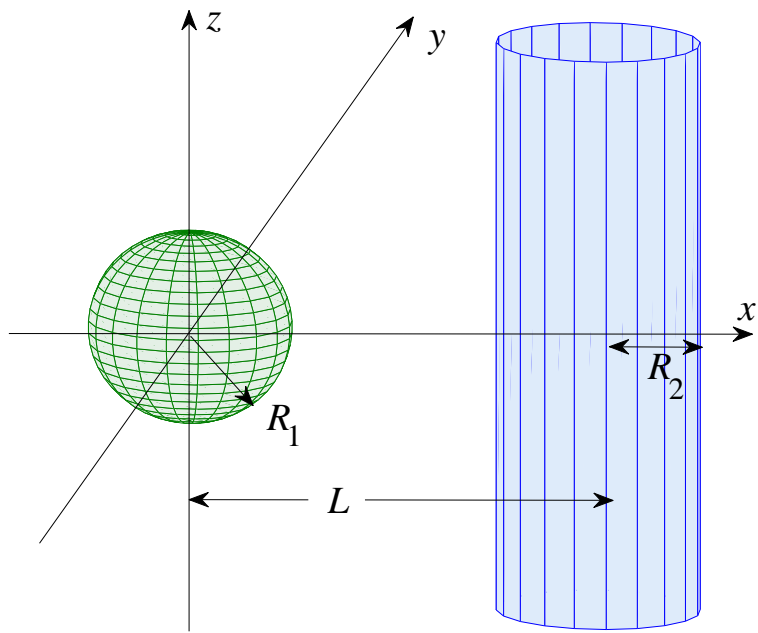} \caption{\label{f1} The configuration of a sphere and a cylinder outside each other.}\end{figure}

In this section, we derive the exact scalar Casimir interaction energy between a sphere and a cylinder. As shown in Fig. \ref{f1}, the centers of the sphere and the cylinder are situated at $(0,0,0)$ and $(L,0,0)$ respectively. The radii of the sphere and the cylinder are given respectively by $R_1$ and $R_2$. The length of the cylinder is denoted by $H$, and we assume that $H\gg R_1, R_2$.

In spherical coordinates $(r,\theta,\phi)$ with center at $\mathbf{O}=(0,0,0)$,
the scalar field $\varphi(\mathbf{x},t)$, $\mathbf{x}=(x,y,z)$,  where $x= r\sin\theta\cos\phi$, $y=r\sin\theta\sin\phi$, $z=r\cos\theta$, can be expanded as
$$\varphi(\mathbf{x},t)=\int_{-\infty}^{\infty} d\omega\sum_{l=0}^{\infty}\sum_{m=-l}^l \left(a_{lm}\varphi_{lm}^{\text{reg}}(\mathbf{x}, k)
+b_{lm}\varphi_{lm}^{\text{out}}(\mathbf{x}, k)\right)e^{-i\omega t}.$$ Here
\begin{equation}\label{eq3_20_10}\begin{split}
&k=\frac{\omega}{c},\\
&\varphi_{l m}^{\text{reg}}(\mathbf{x},k) =\mathcal{C}_{l}^{\text{reg}} j_l(kr)Y_{lm}(\theta,\phi),\quad \mathcal{C}_{l}^{\text{reg}}=i^{-l},\\
&\varphi_{l m}^{\text{out}}(\mathbf{x},k)=
\mathcal{C}_{l}^{\text{out}} h_l^{(1)}(kr)Y_{lm}(\theta,\phi),\quad \mathcal{C}_{l}^{\text{out}}=\frac{\pi}{2}i^{l+2},\\
&Y_{lm}(\theta,\phi)=\sqrt{\frac{2l+1}{4\pi}\frac{(l-m)!}{(l+m)!}}P_l^m(\cos\theta)e^{im\phi},
\end{split}\end{equation}where
 $$j_l(z)=\sqrt{\frac{\pi}{2z}}J_{l+\frac{1}{2}}(z),
\hspace{1cm} h_l^{(1)}(z)=\sqrt{\frac{\pi}{2z}}H_{l+\frac{1}{2}}^{(1)}(z)$$ are the spherical Bessel functions, $Y_{lm}(\theta,\phi)$ are the spherical harmonics and $P_l^m(z)$ are the associated Legendre functions. The constants $\mathcal{C}_{l}^{\text{reg}}$ and $\mathcal{C}_{l}^{\text{out}}$ are chosen so that
\begin{equation*}
\mathcal{C}_{l}^{\text{reg}} j_l(i\zeta) =\sqrt{\frac{\pi}{2\zeta}}I_{l+\frac{1}{2}}(\zeta),\quad \mathcal{C}_{l}^{\text{out}} h_l^{(1)}(i\zeta) =\sqrt{\frac{\pi}{2\zeta}}K_{l+\frac{1}{2}}(\zeta).
\end{equation*}

On the other hand,
the scalar field $\varphi(\mathbf{x},t)$    can also be expanded in cylindrical coordinates $(\rho,\phi,z)$ with center at $\mathbf{O}'=(L,0,0)$:
\begin{equation}\label{eq3_25_2}
\varphi(\mathbf{x},t)=H\int_{-\infty}^{\infty}d\omega \int_{-\infty}^{\infty}\frac{dk_z}{2\pi}\sum_{n=-\infty}^{\infty}\left(
a_{nk_z}\varphi_{nk_z}^{\text{reg}}(\mathbf{x}',k)+b_{nk_z}\varphi_{nk_z}^{\text{out}}(\mathbf{x}',k)\right)e^{-i\omega t},
\end{equation}
where $\mathbf{x}=\mathbf{x}'+\mathbf{O}'$,  $x= x'+L=\rho\cos\phi+L$, $y=y'=\rho\sin\phi$, $z=z'$,
\begin{equation*}
\begin{split}
\varphi_{nk_z}^{\text{reg}}(\mathbf{x}',k)=&\mathcal{C}_{n }^{\text{reg}}J_n(k_{\perp}\rho)e^{in\phi+ik_zz},\quad \mathcal{C}_n^{\text{reg}}= i^{-n}\\
\varphi_{nk_z}^{\text{out}}(\mathbf{x}',k)=&\mathcal{C}_{n }^{\text{out}}H_n^{(1)}(k_{\perp}\rho)e^{in\phi+ik_zz},\quad \mathcal{C}_n^{\text{out}}= \frac{\pi  }{2}i^{n+1}
,\\
k_{\perp}=&\sqrt{k^2-k_z^2}.\\
\end{split}
\end{equation*}
The constants $\mathcal{C}_n^{\text{reg}}$ and $\mathcal{C}_n^{\text{out}}$ are chosen so that
\begin{equation*}
\mathcal{C}_n^{\text{reg}}J_n(i\zeta)=I_n(\zeta),\hspace{1cm}\mathcal{C}_n^{\text{out}}H_n^{(1)}(i\zeta)=K_n(\zeta).
\end{equation*}

 Using multiple scattering formalism \cite{21} or mode summation approach \cite{37}, it was shown that the scalar Casimir interaction energy can be written in the form
\begin{equation}\label{eq3_20_7}\begin{split}
E_{\text{Cas}}
=&\frac{\hbar}{2\pi}\int_0^{\infty} d\xi \text{Tr}\,\ln  \left(\mathbb{I}-\mathbb{M}(i\xi)\right),
\end{split}
\end{equation}
where
$$\mathbb{M}(i\xi)=\mathbb{T}^1(i\xi)\mathbb{U}^{12}(i\xi)\mathbb{T}^2(i\xi)\mathbb{U}^{21}(i\xi).$$
\eqref{eq3_20_7} is called the TGTG formula. The matrices $\mathbb{T}_1$ and $\mathbb{T}_2$ are obtained by matching the boundary conditions on the sphere and on the cylinder respectively. These matrices are related to  scattering matrices \cite{15,21} and have been computed in a lots of literature. They are diagonal matrices. For Dirichlet boundary conditions, the diagonal elements $T^1_{lm}$ and $T^2_{nkz}$ are given by
\begin{equation*}
\begin{split}
 T^1_{lm}(i\xi)=&\frac{I_{l+\frac{1}{2}}\left(R_1\kappa \right)}{K_{l+\frac{1}{2}}\left( R_1\kappa\right)},\\
T^2_{nk_z}(i\xi)=&\frac{I_{n}\left( R_2\sqrt{\kappa^2+k_z^2}\right)}{K_{n}\left( R_2\sqrt{\kappa^2+k_z^2} \right)}.
\end{split}
\end{equation*}For Neumann boundary conditions,
\begin{equation*}
\begin{split}
T^1_{lm}(i\xi)=& \frac{-\frac{1}{2}I_{l+\frac{1}{2}}\left( R_1\kappa\right)
+ R_1 \kappa I_{l+\frac{1}{2}}'\left( R_1\kappa\right)}{-\frac{1}{2}K_{l+\frac{1}{2}}\left( R_1\kappa\right)+ R_1 \kappa K_{l+\frac{1}{2}}'\left( R_1\kappa\right)},\\
 T^2_{nk_z}(i\xi)=& \frac{I_{n}'\left(R_2\sqrt{\kappa^2+k_z^2} \right)}{K_{n}'\left(R_2\sqrt{\kappa^2+k_z^2} \right)}.
\end{split}
\end{equation*}Here
$$\kappa=\frac{\xi}{c}.$$

The hardest part of the problem in is to compute the translation matrices $\mathbb{U}^{12}$ and $\mathbb{U}^{21}$. This has not been worked out before. In the following, we generalize the operator approach introduced by Wittman \cite{36} (see also \cite{37}) to calculate these translation matrices.

As in \cite{36,37}, we introduce the differential operator $\mathcal{P}_{lm}$ defined by
\begin{equation*}
\begin{split}
\mathcal{P}_{lm}=&(-1)^m \sqrt{\frac{2l+1}{4\pi}\frac{(l-m)!}{(l+m)!}}\left(\frac{\pa_x+i\pa_y}{ik}\right)^mP_l^{(m)}\left(\frac{\pa_z}{ik}\right), \\
\mathcal{P}_{l,-m}=&  \sqrt{\frac{2l+1}{4\pi}\frac{(l-m)!}{(l+m)!}}\left(\frac{\pa_x-i\pa_y}{ik}\right)^mP_l^{(m)}\left(\frac{\pa_z}{ik}\right).
\end{split}
\end{equation*}Here $m\geq 0$,  $P_l(z)$ is the Legendre polynomial of degree $l$ and $P_l^{(m)}(z)$ is its $m$-times derivative. It follows from the definition of
spherical harmonics \cite{38} that applying to $e^{i\mathbf{k}\cdot\mathbf{r}}$, where $\mathbf{k}=k_x\mathbf{e}_x+\mathbf{k}_y\mathbf{e}_y+\mathbf{k_z}e_z $, $\mathbf{r}=x\mathbf{e}_x+y\mathbf{e}_y+z\mathbf{e}_z$, we have
\begin{equation}\label{eq12_4_1}
\mathcal{P}_{lm}e^{i\mathbf{k}\cdot\mathbf{r}}=Y_{lm}(\theta_k,\phi_k)e^{i\mathbf{k}\cdot\mathbf{r}}
\end{equation}for $l=0,1,2,\ldots$, $-l\leq m\leq l$. On the other hand, one can show by induction that
\begin{equation}\label{eq5_2_1}
\begin{split}
\mathcal{P}_{lm}j_0(kr)=i^lj_l(kr)Y_{lm}(\theta,\phi),\hspace{1cm}\mathcal{P}_{lm}h_0^{(1)}(kr)=i^lh_l^{(1)}(kr)Y_{lm}(\theta,\phi).
\end{split}
\end{equation}
It can then be shown that the spherical wave functions $\varphi_{lm}^{\text{reg}}(\mathbf{x},k)$  and $\varphi_{lm}^{\text{out}}(\mathbf{x},k)$ have the following integral representations (see \cite{36,37}):\begin{equation}\label{eq5_9_1}
\begin{split}
\varphi_{lm}^{\text{reg}}(\mathbf{x},k)
=&\frac{1}{4\pi i^l}\int_0^{2\pi}d\phi_k\int_0^{\pi}d\theta_k \sin\theta_k Y_{lm}(\theta_k,\phi_k)e^{i\mathbf{k}\cdot\mathbf{r}},\\
\varphi_{lm}^{\text{out}}(\mathbf{x},k)
=&\frac{\mathcal{C}_l^{\text{out}}}{2\pi i^l}
\int_{-\infty}^{\infty} dk_x\int_{-\infty}^{\infty} dk_y  Y_{lm}(\theta_k,\phi_k)\frac{e^{ik_xx+ik_yy \pm i\sqrt{k^2-k_x^2-k_y^2}z}}{k\sqrt{k^2-k_x^2-k_y^2}},\quad z\gtrless 0.
\end{split}
\end{equation}

For the cylindrical wave functions $\varphi_{nk_z}^{\text{reg}}(\mathbf{x}',k)$ and $\varphi_{nk_z}^{\text{out}}(\mathbf{x}',k)$, we introduce the operator $\mathcal{Q}_n$ as in \cite{37}. For $n\geq 0$,
\begin{equation*}
\begin{split}
\mathcal{Q}_n=&\left(\frac{\pa_{x'}+i\pa_{y'}}{ik_{\perp}}\right)^n, \\
\mathcal{Q}_{-n}=&\left(\frac{\pa_{x'}-i\pa_{y'}}{ik_{\perp}}\right)^n.
\end{split}
\end{equation*} It follows that
\begin{equation*}
\mathcal{Q}_ne^{i\mathbf{k}_{\perp}\cdot\boldsymbol{\rho}}=e^{in\phi_k}e^{i\mathbf{k}_{\perp}\cdot\boldsymbol{\rho}}
\end{equation*}for all $n$, where $\mathbf{k}_{\perp}=k_x\mathbf{e}_x+k_y\mathbf{e}_y$, $\boldsymbol{\rho}=x'\mathbf{e}_x+y'\mathbf{e}_y$.
One can then show that (see \cite{37}):
\begin{equation}\label{eq4_10_5}
\begin{split}
 \varphi_{nk_z}^{\text{reg}}(\mathbf{x}',k)
=&\frac{\mathcal{C}_n^{\text{reg}}}{2\pi i^n} \int_0^{2\pi}d\phi_k e^{in\phi_k}
 e^{i\mathbf{k}_{\perp}\cdot\boldsymbol{\rho}+ik_zz'},\\
 \varphi_{nk_z}^{\text{out}}(\mathbf{x}',k) =& \frac{\mathcal{C}_n^{\text{out}}}{ \pi i^n} \int_{-\infty}^{\infty}dk_y e^{in\phi_k}
\frac{e^{\pm i\sqrt{k_{\perp}^2-k_y^2}x'+ik_yy'+ik_zz'}}{\sqrt{k_{\perp}^2-k_y^2}},\quad x\gtrless 0.
\end{split}\end{equation}

The translation matrices $\mathbb{U}^{12}$ and $\mathbb{U}^{21}$ are defined so that their elements $U^{12}_{lm,nk_z} $ and $U^{21}_{nk_z,lm}$ satisfy the following identities:
 \begin{equation}\label{eq4_10_2}\begin{split}
\varphi_{nk_z}^{\text{out}}(\mathbf{x}-\mathbf{O}',k)=&\sum_{l=0}^{\infty}\sum_{m=-l}^{l} U^{12}_{lm,nk_z} \varphi_{lm}^{\text{reg}}(\mathbf{x},k),\\
\varphi_{lm}^{\text{out}}(\mathbf{x}'+\mathbf{O}',k)=&\sum_{n=-\infty}^{\infty}H\int_{-\infty}^{\infty}\frac{dk_z}{2\pi}U^{21}_{nk_z,lm}
\varphi_{nk_z}^{\text{reg}}(\mathbf{x}',k).\end{split}\end{equation}
To find $U^{12}_{lm,nk_z}$, notice that \eqref{eq12_4_1} and \eqref{eq5_9_1} imply that
\begin{equation}\label{eq4_9_6}
\begin{split}
\left(P_{l''m''}\varphi_{lm}^{\text{reg}}\right)(\mathbf{0})=&\frac{\mathcal{C}_{l}^{\text{reg}}}{4\pi i^{l}}\int_0^{2\pi}d\phi_k\int_0^{\pi}d\theta_k \sin\theta_k
Y_{lm}(\theta_k,\phi_k)
Y_{l''m''}(\theta_k,\phi_k)=\frac{(-1)^{m}}{4\pi i^{l}}\delta_{l,l''}\delta_{m'',-m}\mathcal{C}_{l}^{\text{reg}}.
\end{split}\end{equation}Therefore, applying $\mathcal{P}_{l,-m}$ to both sides of the first equation of \eqref{eq4_10_2} and set $\mathbf{x}=\mathbf{0}$, we have
\begin{equation*}
U^{12}_{lm,nk_z} =\frac{4\pi i^{l}}{\mathcal{C}_{l}^{\text{reg}}}(-1)^{m}\left(\mathcal{P}_{l,-m}\varphi_{nk_z}^{\text{out}}\right)( -\mathbf{O}',k).\end{equation*}Using \eqref{eq4_10_5} and \eqref{eq12_4_1}, we then find that \begin{equation*}
\begin{split}
U^{12}_{lm,nk_z} =& 2 \pi(-1)^{l+m} i\int_{-\infty}^{\infty}dk_y   Y_{l,-m}(\theta_k,\phi_k)
e^{in\phi_k}\frac{e^{i\sqrt{k_{\perp}^2-k_y^2}L}}{\sqrt{k_{\perp}^2-k_y^2}}\\
=&\sqrt{4\pi(2l+1)\frac{(l-m)!}{(l+m)!}}(-1)^{l+m+n}P_l^m\left(\frac{k_z}{k}\right)\varphi_{n-m,k_z}^{\text{out}}\left(\mathbf{O}',k\right).
\end{split}
\end{equation*}
Passing to imaginary frequency with $k=i\kappa$ gives
\begin{equation}\label{eq12_4_2}
U^{12}_{lm,nk_z}(i\xi) =\sqrt{4\pi(2l+1)\frac{(l-m)!}{(l+m)!}}(-1)^{n}P_l^m\left(\frac{ik_z}{\kappa}\right)K_{n-m}\left(L\sqrt{\kappa^2+k_z^2}  \right).\end{equation}

For $U^{21}_{nk_z,lm}$, let us consider the case where $m\geq 0$. The case where $m<0$ is analogous.  Using \eqref{eq5_2_1} and \eqref{eq4_10_5},  the second equation of \eqref{eq4_10_2} can be written as
\begin{equation*}
\begin{split}
& \mathcal{C}_l^{\text{out}}  i^{-l}\mathcal{P}_{lm}h_0^{(1)}\left(k|\mathbf{x}'+\mathbf{O}'|\right)\\
=&\sum_{n=-\infty}^{\infty}H\int_{-\infty}^{\infty}\frac{dk_z}{2\pi}U^{21}_{nk_z,lm}
\frac{\mathcal{C}_n^{\text{reg}}}{2\pi i^n}\int_0^{2\pi}d\phi_k e^{in\phi_k} e^{i\sqrt{k^2-k_z^2}\cos\phi_kx'+i\sqrt{k^2-k_z^2}\sin\phi_ky'+ik_zz'}.
\end{split}\end{equation*}
In imaginary frequency, we have
\begin{equation*}
\begin{split}
&\sum_{n=-\infty}^{\infty}H\int_{-\infty}^{\infty}\frac{dk_z}{2\pi}U_{nk_z,lm}^{21}
( i\xi)\frac{(-1)^n}{2\pi  }\int_0^{2\pi}d\phi_k e^{in\phi_k} e^{-\sqrt{\kappa^2+k_z^2}\cos\phi_kx'-\sqrt{\kappa^2+k_z^2}\sin\phi_ky'+ik_zz'}\\=&
 \frac{\pi}{2}\sqrt{\frac{2l+1}{4\pi}\frac{(l-m)!}{(l+m)!}}\left(\frac{\pa_{x'}+i\pa_{y'}}{\kappa}\right)^m P_l^{(m)}\left(-\frac{\pa_{z'}}{\kappa}\right)
 \frac{e^{-\kappa|\mathbf{x}'+\mathbf{O}'|}}{\kappa|\mathbf{x}'+\mathbf{O}'|}.
\end{split}\end{equation*}
The left hand side is an inverse Fourier transform. Taking Fourier transform, we find that
\begin{equation*}
\begin{split}
&\sum_{n=-\infty}^{\infty} U_{nk_z,lm}^{21}
(i\xi)\frac{1}{2\pi  }\int_0^{2\pi}d\phi_k e^{in\phi_k} e^{-\sqrt{\kappa^2+k_z^2}\cos\phi_kx'-\sqrt{\kappa^2+k_z^2}\sin\phi_ky' }\\=&
 \frac{\pi(-1)^n}{2 H}\sqrt{\frac{2l+1}{4\pi}\frac{(l-m)!}{(l+m)!}}\left(\frac{\pa_{x'}+i\pa_{y'}}{\kappa}\right)^m \int_{-\infty}^{\infty} dz' \left[P_l^{(m)}\left(-\frac{\pa_{z'}}{\kappa}\right)
 \frac{e^{-\kappa|\mathbf{x}'+\mathbf{O}'|}}{\kappa|\mathbf{x}'+\mathbf{O}'|}\right]e^{-ik_zz'}.
\end{split}\end{equation*}
Applying integration by parts, we find that
\begin{equation*}
\begin{split}
&\int_{-\infty}^{\infty} dz' \left[P_l^{(m)}\left(-\frac{\pa_{z'}}{\kappa}\right)
 \frac{e^{-\kappa|\mathbf{x}'+\mathbf{O}'|}}{\kappa|\mathbf{x}'+\mathbf{O}'|}\right]e^{-ik_zz'}\\
=&\int_{-\infty}^{\infty} dz'
 \frac{e^{-\kappa|\mathbf{x}'+\mathbf{O}'|}}{\kappa|\mathbf{x}'+\mathbf{O}'|}\left[P_l^{(m)}\left(\frac{\pa_{z'}}{\kappa}\right)e^{-ik_zz'}\right]\\
 =&P_l^{(m)}\left(-\frac{ik_z}{\kappa}\right)\int_{-\infty}^{\infty} dz'
 \frac{e^{-\kappa\sqrt{(x'+L)^2+y^{\prime 2}+z^{\prime 2}}}}{\kappa\sqrt{(x'+L)^2+y^{\prime 2}+z^{\prime 2}}}e^{-ik_zz'}\\
 =&(-1)^m\left(\frac{\kappa}{\sqrt{\kappa^2+k_z^2}}\right)^mP_l^m\left(-\frac{ik_z}{\kappa}\right)\frac{2}{\kappa}
 K_0\left(\sqrt{\kappa^2+k_z^2}\sqrt{(x'+L)^2+y^{\prime 2}}\right)\\
  =&(-1)^m\left(\frac{\kappa}{\sqrt{\kappa^2+k_z^2}}\right)^mP_l^m\left(-\frac{ik_z}{\kappa}\right)\frac{1}{\kappa}
 \int_{-\infty}^{\infty}dk_y\frac{e^{- \sqrt{\kappa^2+k_z^2+k_y^2}(x'+L)+ik_yy'}}{\sqrt{\kappa^2+k_z^2+k_y^2}}.
\end{split}
\end{equation*}
Hence,
\begin{equation*}
\begin{split}
&\sum_{n=-\infty}^{\infty} U_{nk_z,lm}^{21}
(i\xi)\frac{1}{2\pi  }\int_0^{2\pi}d\phi_k e^{in\phi_k} e^{-\sqrt{\kappa^2+k_z^2}\cos\phi_kx'-\sqrt{\kappa^2+k_z^2}\sin\phi_ky' }\\=&
 \frac{\pi(-1)^{n }}{ 2 H\kappa}\sqrt{\frac{2l+1}{4\pi}\frac{(l-m)!}{(l+m)!}} P_l^m\left(-\frac{ik_z}{\kappa}\right)\mathcal{Q}_m
\int_{-\infty}^{\infty}dk_y\frac{e^{- \sqrt{\kappa^2+k_z^2+k_y^2}(x'+L)+ik_yy'}}{\sqrt{\kappa^2+k_z^2+k_y^2}}.
\end{split}\end{equation*}
Applying the operator
$\mathcal{Q}_{-n}$ to both sides and setting $x'=y'=0$, we find that
\begin{equation}\label{eq4_10_6}
\begin{split}
   U_{nk_z,lm}^{21}
(i\xi)  =&
 \frac{\pi(-1)^{n }}{ 2 H\kappa}\sqrt{\frac{2l+1}{4\pi}\frac{(l-m)!}{(l+m)!}} P_l^m\left(-\frac{ik_z}{\kappa}\right)
\int_{-\infty}^{\infty}dk_y e^{i(m-n)\phi_k}\frac{e^{- \sqrt{\kappa^2+k_z^2+k_y^2}L}}{\sqrt{\kappa^2+k_z^2+k_y^2}}\\
 =&\frac{\pi(-1)^{n }}{ H\kappa}\sqrt{\frac{2l+1}{4\pi}\frac{(l-m)!}{(l+m)!}} P_l^m\left(-\frac{ik_z}{\kappa}\right)K_{m-n}(L\sqrt{\kappa^2+k_z^2}).
\end{split}
\end{equation}

Now we can write down the formula for  the scalar Casimir interaction energy between a sphere and a cylinder. It is given by \eqref{eq3_20_7},
where the trace Tr is
\begin{equation*}
\text{Tr}=\sum_{l=0}^{\infty}\sum_{m=-l}^{l},
\end{equation*}
and
\begin{equation*}
\begin{split}
M_{lm,l'm'}(i\xi)=&\frac{T^1_{lm}}{2\kappa}\sqrt{ (2l+1)(2l'+1)\frac{(l-m)!}{(l+m)!}\frac{(l'-m')!}{(l'+m')!}}\\&\times\sum_{n=-\infty}^{\infty}\int_{-\infty}^{\infty} dk_z
P_l^m\left(\frac{ik_z}{\kappa}\right)K_{n-m}\left(L\sqrt{\kappa^2+k_z^2} \right)T^2_{nk_z}
 P_{l'}^{m'}\left(-\frac{ik_z}{\kappa}\right)K_{m'-n}\left(L\sqrt{\kappa^2+k_z^2} \right).
\end{split}\end{equation*}Making a change of variables
$$k_z=\kappa \sinh\theta,$$ we have
\begin{equation}\label{eq12_10_1}
\begin{split}
M_{lm,l'm'}(i\xi)=&\frac{T^1_{lm}}{2 }\sqrt{ (2l+1)(2l'+1)\frac{(l-m)!}{(l+m)!}\frac{(l'-m')!}{(l'+m')!}}\\&\times\sum_{n=-\infty}^{\infty}\int_{-\infty}^{\infty} d\theta \cosh\theta
P_l^m\left(i\sinh\theta\right)K_{n-m}\left(L\kappa \cosh\theta\right)T^2_{nk_z}
 P_{l'}^{m'}\left(-i\sinh\theta\right)K_{m'-n}\left(L\kappa \cosh\theta\right).
\end{split}\end{equation}

The formula  for $M_{lm,l'm'}(i\xi)$ \eqref{eq12_10_1} contains $P_{l}^m(\pm i\sinh\theta)$ and  $P_{l}^{m\prime}(\pm i\sinh\theta)$ which are complex. This leads to some doubt whether the Casimir interaction energy \eqref{eq3_20_7} is real. In the following, we show that this is indeed the case. Moreover, the Casimir energy is always negative.

Expanding the logarithm in \eqref{eq3_20_7}, we have
\begin{equation}\label{eq12_5_6}\begin{split}
E_{\text{Cas}}
=&-\sum_{s=1}^{\infty}\frac{1}{s}\frac{\hbar}{2\pi}\int_0^{\infty} d\xi \prod_{i=1}^s\left(\sum_{l_i=0}^{\infty}\sum_{m_i=-l_i}^{l_i} \right)\prod_{i=1}^sM_{l_im_i,l_{i+1}m_{i+1}}(i\xi),
\end{split}
\end{equation}with the convention that $l_{s+1}=l_1$ and $m_{s+1}=m_1$.

Now for any  $\nu$ and for any positive real number $z$,  $I_{\nu}(z), I_{\nu}'(z)$ and $K_{\nu}(z)$ are positive whereas $K_{\nu}'(z)$ is negative. Moreover, for $l\geq 0$,
\begin{equation*}
-\frac{1}{2}I_{l+\frac{1}{2}}(z)+zI_{l+\frac{1}{2}}'(z)>0,\hspace{1cm}-\frac{1}{2}K_{l+\frac{1}{2}}(z)+zK_{l+\frac{1}{2}}'(z)<0.
\end{equation*}Hence, we find that
for either Dirichlet boundary conditions or Neumann boundary conditions,
$$T^1_{lm}T^2_{nk_z}>0.$$
Using the representation \eqref{eq12_10_1}, we find that
\begin{equation}\label{eq12_5_8}
\prod_{i=1}^sM_{l_im_i,l_{i+1}m_{i+1}}(i\xi)
\end{equation}can be written as
\begin{equation*}\begin{split}
\prod_{i=1}^sM_{l_im_i,l_{i+1}m_{i+1}}(i\xi)=&\left(\prod_{i=1}^s\int_{-\infty}^{\infty}d\theta_i\right) \left(\text{positive function}\times \prod_{i=1}^s P_{l_i}^{m_i}(i\sinh\theta_i)P_{l_{i+1}}^{m_{i+1}}(-i\sinh\theta_i)\right).\end{split}
\end{equation*}
Moreover, the positive function is also even in each of the variables $\theta_i$.

 Using the formula (see \cite{38}) for associated Legendre functions:
\begin{equation*}\begin{split}
P_l^m(z)=&\frac{(-1)^m}{2^ll!}(1-z^2)^{m/2}\frac{d^l}{dz^l}(z^2-1)^l,\\
P_l^{-m}(z)=&(-1)^m\frac{(l-m)!}{(l+m)!}P_l^m(z),
\end{split}\quad m\geq 0,\end{equation*}we find that for $m\geq 0$,
\begin{equation*}
\begin{split}
P_l^m(\pm i\sinh\theta)=&(\pm i)^{l+m}\frac{1}{2^ll!}\cosh^m\theta \sum_{j=0}^{\left[\frac{l-m}{2}\right]}  \begin{pmatrix} l\\j\end{pmatrix} \frac{ (2l-2j)!}{(l-2j-m)!}\sinh^{l-2j-m}\theta.
\end{split}
\end{equation*}
Notice that except for the factor $(\pm i)^{l+m}$, the rest are real. Hence, in the product $$\prod_{i=1}^s P_{l_i}^{m_i}(i\sinh\theta_i)P_{l_{i+1}}^{m_{i+1}}(-i\sinh\theta_i),$$
the factor $i^{l_i+m_i-l_{i+1}-m_{i+1}}$   will cancel off. This shows that \eqref{eq12_5_8} is real. For integration over $\theta_i$, we will have a  function  that is positive and even in $\theta_i$ multiply with
\begin{equation}\label{eq12_5_5}(\sinh\theta_i)^{l_i-m_i+l_{i+1}-m_{i+1}}.\end{equation} This function is positive and even if and only if $\left(l_i-m_i+l_{i+1}-m_{i+1}\right)$ is even, or in other words, $\left(l_i-m_i\right)$ has the same parity as $\left(l_{i+1}-m_{i+1}\right)$. When $\left(l_i-m_i+l_{i+1}-m_{i+1}\right)$ is odd, \eqref{eq12_5_5} is an odd function in $\theta_i$ and therefore the integration over $\theta_i$ vanishes. In summary, the function \eqref{eq12_5_8} is positive if all $\left(l_i-m_i\right)$, $i=1,\ldots, s,$ have the same parity. Otherwise, it is zero. Hence, the representation \eqref{eq12_5_6} shows that the Casimir interaction energy is always negative.

For the Casimir interaction force $F_{\text{Cas}}$, it is defined as the negative of the derivative of the Casimir interaction energy with respect to $d$, the distance between the sphere and the cylinder. Namely,
$$F_{\text{Cas}}=-\frac{\partial}{\partial d}E_{\text{Cas}}.$$
Notice that $d$ only appears in $L$: $$L=R_1+R_2+d,$$and in \eqref{eq12_10_1}, $L$ only appears in the two terms $K_{n-m}\left(L\kappa \cosh\theta\right)$ and
 $K_{m'-n}\left(L\kappa \cosh\theta\right)$. In differentiating \eqref{eq12_5_8} with respect to $d$, we will have $2s$ terms, each one is obtained by differentiating one of these modified Bessel functions which give a negative function. Combining with the same argument as before, we see that the Casimir interaction force is always negative. In other words, the force is attractive. This is true for all distances.

   \section{Exact electromagnetic Casimir interaction energy}\label{sec3}

In this section, we consider electromagnetic Casimir interaction between a perfectly conducting sphere and a perfectly conducting cylinder. The electric field $\mathbf{E}$ and the magnetic field $\mathbf{B}$ can be expressed in terms of the vector potential $\mathbf{A}$ by
\begin{equation*}
\mathbf{E}=-\frac{\pa\mathbf{A}}{\pa t},\hspace{1cm}\mathbf{B}=\nabla\times \mathbf{A}.
\end{equation*}

In spherical coordinates centered at $(0,0,0)$, the vector potential $\mathbf{A}(\mathbf{x},t)$ can be expanded as
\begin{equation}\label{eq3_22_12}
\mathbf{A}(\mathbf{x},t)=\int_{-\infty}^{\infty} d\omega\sum_{l=1}^{\infty}\sum_{m=-l}^l \left(a_{lm}\mathbf{A}_{lm}^{\text{TE, reg}}(\mathbf{x}, k)
+b_{lm}\mathbf{A}_{lm}^{\text{TE, out}}(\mathbf{x}, k)
+c_{lm}\mathbf{A}_{lm}^{\text{TM, reg}}(\mathbf{x}, k)+d_{lm}\mathbf{A}_{lm}^{\text{TM, out}}(\mathbf{x}, k)\right)e^{-i\omega t},
\end{equation}where
\begin{equation*}\begin{split}
\mathbf{A}_{lm}^{\text{TE},*}(\mathbf{x},k)=&\frac{i}{\sqrt{l(l+1)}} \nabla\times \varphi_{l m}^{*}(\mathbf{x},k) r\mathbf{e}_r \\
=&  \mathcal{C}_{l}^* f_l^*(kr)\mathbf{X}_{lm}(\theta,\phi),\\
\mathbf{A}_{lm}^{\text{TM},*}(\mathbf{x},k)=&\frac{i}{k\sqrt{l(l+1)}}\nabla\times \nabla\times \varphi_{l m}^{*}(\mathbf{x},k) r\mathbf{e}_r.
\end{split}\end{equation*}Here $*=$ reg or out, with $f_l^{\text{reg}}(z)=j_l(z)$ and $f_l^{\text{out}}(z)=h_l^{(1)}(z)$,
$$\mathbf{X}_{lm}(\theta,\phi)=\frac{1}{i\sqrt{l(l+1)}}\mathbf{r}\times\nabla Y_{lm}(\theta,\phi)=-\frac{m}{\sin\theta}Y_{lm}(\theta,\phi)\mathbf{e}_{\theta}-i\frac{\pa Y_{lm}(\theta,\phi)}{\pa\theta}
\mathbf{e}_{\phi}$$ are vector spherical harmonics \cite{38}.
Obviously,
\begin{equation}\label{eq12_4_3}
\mathbf{A}_{lm}^{\text{TM},*}(\mathbf{x},k)=\frac{1}{k}\nabla\times \mathbf{A}_{lm}^{\text{TE},*}(\mathbf{x},k).
\end{equation}Straightforward computation gives
\begin{equation}\label{eq12_4_4}
\mathbf{A}_{lm}^{\text{TE},*}(\mathbf{x},k)=\frac{1}{k}\nabla\times \mathbf{A}_{lm}^{\text{TM},*}(\mathbf{x},k).
\end{equation}

In cylindrical coordinates centered at $(L,0,0)$,  $\mathbf{A}(\mathbf{x},t)$ can be expanded as
\begin{equation}\label{eq12_10_3}\begin{split}
\mathbf{A}(\mathbf{x},t)=&H\int_{-\infty}^{\infty}d\omega \int_{-\infty}^{\infty}\frac{dk_z}{2\pi}\sum_{n=-\infty}^{\infty}\left(a_{nk_z}\mathbf{A}_{nk_z}^{\text{TE, reg}}(\mathbf{x}', k)
+b_{nk_z}\mathbf{A}_{nk_z}^{\text{TE, out}}(\mathbf{x}', k)\right.\\&\hspace{4cm}\left.
+c_{nk_z}\mathbf{A}_{nk_z}^{\text{TM, reg}}(\mathbf{x}', k)+d_{nk_z}\mathbf{A}_{nk_z}^{\text{TM, out}}(\mathbf{x}', k)\right)e^{-i\omega t},\end{split}
\end{equation}
where\begin{equation*}
\begin{split}
\mathbf{A}_{nk_z}^{\text{TE}, *}(\mathbf{x}',k)=&\frac{1}{k_{\perp}}\nabla\times \varphi_{nk_z}^*\mathbf{e}_z\\
=& \mathcal{C}_n^*\left(\frac{in}{k_{\perp}\rho }g_n^*(k_{\perp}\rho)\mathbf{e}_{\rho}-g_n^{*\prime}(k_{\perp}\rho)\mathbf{e}_{\phi}\right)e^{in\phi+ik_zz},\\
\mathbf{A}_{nk_z}^{\text{TM}, *}(\mathbf{x}',k)=&\frac{1}{kk_{\perp}}\nabla\times\nabla\times \varphi_{nk_z}^*\mathbf{e}_z\\
=&\mathcal{C}_n^*\left(\frac{ik_z}{k}g_n^{*\prime}(k_{\perp}\rho)\mathbf{e}_{\rho}-\frac{nk_z}{kk_{\perp}\rho}g_n^*(k_{\perp}\rho)\mathbf{e}_{\phi}+\frac{k_{\perp}}{k}
g_n^*(k_{\perp}\rho)\mathbf{e}_z\right)e^{in\phi+ik_zz}.
\end{split}
\end{equation*}Here $g_n^{\text{reg}}(z)=J_n(z)$ and $g_n^{\text{out}}(z)=H_n^{(1)}(z)$. We have
\begin{equation}\label{eq12_4_5}\begin{split}
\mathbf{A}_{nk_z}^{\text{TM},*}(\mathbf{x},k)=&\frac{1}{k}\nabla\times \mathbf{A}_{nk_z}^{\text{TE},*}(\mathbf{x},k),\\
\mathbf{A}_{nk_z}^{\text{TE},*}(\mathbf{x},k)=&\frac{1}{k}\nabla\times \mathbf{A}_{nk_z}^{\text{TM},*}(\mathbf{x},k).
\end{split}
\end{equation}

Using multiple scattering approach \cite{24} or mode summation approach \cite{37}, one still finds that the Casimir interaction energy can be written in the form \eqref{eq3_20_7}, but the components of the matrices $\mathbb{T}^1, \mathbb{U}^{12}, \mathbb{T}^2, \mathbb{U}^{21}$ are $2\times 2$ matrices. For perfectly conducting sphere and cylinder, the matrices $\mathbb{T}^1$ and $\mathbb{T}^2$  are well known. They are diagonal and each $2\times 2$ component is also diagonal, i.e.,
\begin{equation*}
\mathbb{T}^1_{lm}=\begin{pmatrix} T^{1,\text{TE}}_{lm} & 0\\ 0 &T^{1,\text{TM}}_{lm}\end{pmatrix},\quad \mathbb{T}^2_{nk_z}=\begin{pmatrix} T^{2,\text{TE}}_{nk_z} & 0\\ 0 &T^{2,\text{TM}}_{nk_z}\end{pmatrix},
\end{equation*}with
\begin{equation*}
\begin{split}
T^{1,\text{TE}}_{lm}(i\xi)=&\frac{I_{l+\frac{1}{2}}\left(R_1\kappa \right)}{K_{l+\frac{1}{2}}\left( R_1\kappa\right)},\\
T^{1,\text{TM}}_{lm}(i\xi)=& \frac{\frac{1}{2}I_{l+\frac{1}{2}}\left( R_1\kappa\right)
+ R_1 \kappa I_{l+\frac{1}{2}}'\left( R_1\kappa\right)}{\frac{1}{2}K_{l+\frac{1}{2}}\left( R_1\kappa\right)+ R_1 \kappa K_{l+\frac{1}{2}}'\left( R_1\kappa\right)},\\
T^{2,\text{TE}}_{nk_z}(i\xi)=&\frac{I_{n}'\left( R_2\sqrt{\kappa^2+k_z^2}\right)}{K_{n}'\left( R_2\sqrt{\kappa^2+k_z^2} \right)},\\
T^{2,\text{TM}}_{nk_z}(i\xi)=&\frac{I_{n}\left( R_2\sqrt{\kappa^2+k_z^2}\right)}{K_{n}\left( R_2\sqrt{\kappa^2+k_z^2} \right)}.
\end{split}
\end{equation*}

The translation matrices $\mathbb{U}^{12}$ and $\mathbb{U}^{21}$ have not been worked out before. The main results of this section are the explicit formulas for these translation matrices. We use the same approach as in the previous section, which is inspired by \cite{36}. First define the vector-valued operator \cite{36}:
\begin{equation}
\begin{split}
\boldsymbol{\mathcal{P}}_{lm}=&\frac{1}{\sqrt{l(l+1)}}\left(\boldsymbol{\mathbf{L}}\mathcal{P}_{lm}-\mathcal{P}_{lm}\boldsymbol{\mathbf{L}}\right),
\end{split}\end{equation}
where
$$\mathbf{L}=\frac{1}{i}\mathbf{r}\times\nabla.$$
In \cite{36}, it has been shown that
\begin{equation}\label{eq5_3_4}
\begin{split}
\boldsymbol{\mathcal{P}}_{lm}
=&\frac{1}{\sqrt{l(l+1)}}\Biggl(\frac{\mathbf{e}_x}{2}\left[\sqrt{(l-m)(l+m+1)}\mathcal{P}_{l,m+1}+\sqrt{(l+m)(l-m+1)}\mathcal{P}_{l,m-1}\right]\\
&+\frac{\mathbf{e}_y}{2i}\left[\sqrt{(l-m)(l+m+1)}\mathcal{P}_{l,m+1}-\sqrt{(l+m)(l-m+1)}\mathcal{P}_{l,m-1}\right]
+m\mathbf{e}_z\mathcal{P}_{lm}\Biggr),
\end{split}\end{equation}
\begin{equation}\label{eq4_3_9}
\boldsymbol{\mathcal{P}}_{lm}e^{i\mathbf{k}\cdot\mathbf{r}}=\mathbf{X}_{lm}(\theta_k,\phi_k)e^{i\mathbf{k}\cdot\mathbf{r}},
\end{equation}
and
\begin{equation}\label{eq5_3_3}
\mathbf{A}^{\text{TE, reg}}_{lm}(\mathbf{x},k)=\mathcal{C}_l^{\text{reg}}i^{-l}\boldsymbol{\mathcal{P}}_{lm}j_0(kr),\quad \mathbf{A}^{\text{TE, out}}_{lm}=
\mathcal{C}_l^{\text{out}}i^{-l}\boldsymbol{\mathcal{P}}_{lm}h_0(kr).
\end{equation}
These imply that
\begin{equation}\label{eq4_3_10}
\begin{split}
\mathbf{A}^{\text{TE, reg}}_{lm}(\mathbf{x},k)=&\frac{\mathcal{C}_l^{\text{reg}}}{4\pi i^{l}}\int_0^{2\pi}d\phi_k\int_0^{\pi}d\theta_k\sin\theta_k
\mathbf{X}_{lm}(\theta_k,\phi_k)e^{i\mathbf{k}\cdot\mathbf{r}},\\
\mathbf{A}^{\text{TM, reg}}_{lm}(\mathbf{x},k)=&\frac{\mathcal{C}_l^{\text{reg}}}{4\pi i^{l}}
\int_0^{2\pi}d\phi_k\int_0^{\pi} d\theta_k\sin\theta_k\frac{i\mathbf{k}}{k}\times
\mathbf{X}_{lm}(\theta_k,\phi_k)e^{i\mathbf{k}\cdot\mathbf{r}}.
\end{split}
\end{equation}

For the  vector cylindrical waves, consider the operator \cite{37}:
\begin{equation*}
\boldsymbol{\mathcal{Q}}_n=\frac{ \mathbf{e}_x}{2}\left(\mathcal{Q}_{n+1}-\mathcal{Q}_{n-1}\right)+\frac{ \mathbf{e}_y}{2i}\left(\mathcal{Q}_{n+1}+\mathcal{Q}_{n-1}\right).
\end{equation*}
In \cite{37}, it has been shown that
$$\mathbf{A}_{nk_z}^{\text{TE},*}(\mathbf{x}',k) =i^{-n}\mathcal{C}_n^*\boldsymbol{\mathcal{Q}}_ng_0^*(k_{\perp}\rho)e^{ik_zz}.$$
It follows that
\begin{equation}\label{eq4_9_4}
\begin{split}
\mathbf{A}_{nk_z}^{\text{TE, reg}}(\mathbf{x}',k)
=&\frac{\mathcal{C}_n^{\text{reg}}}{2\pi i^n} \int_0^{2\pi}d\phi_k \left( -i \mathbf{e}_{\phi_k}\right)e^{in\phi_k}
e^{i\mathbf{k}_{\perp}\cdot\boldsymbol{\rho}+ik_zz},\\
\mathbf{A}_{nk_z}^{\text{TM, reg}}(\mathbf{x}',k)
=&\frac{\mathcal{C}_n^{\text{reg}}}{2\pi i^n} \int_0^{2\pi}d\phi_k  \left( -\mathbf{e}_{\theta_k}\right)e^{in\phi_k}e^{i\mathbf{k}_{\perp}\cdot\boldsymbol{\rho}+ik_zz},
\end{split}
\end{equation}
\begin{equation}\label{eq4_9_5}
\begin{split}
\mathbf{A}_{nk_z}^{\text{TE, out}}(\mathbf{x}',k)=&\frac{\mathcal{C}_n^{\text{out}}}{ \pi i^n} \int_
{-\infty}^{\infty}dk_y \left( -i\mathbf{e}_{\phi_k}\right)e^{in\phi_k}\frac{e^{\pm i
\sqrt{k_{\perp}^2-k_y^2}x'+ik_yy'+ik_zz'}}{\sqrt{k_{\perp}^2-k_y^2}},\quad x\gtrless 0,\\
\mathbf{A}_{nk_z}^{\text{TM, out}}(\mathbf{x}',k) =&\frac{\mathcal{C}_n^{\text{out}}}{ \pi i^n} \int_
{-\infty}^{\infty}dk_y \left( -\mathbf{e}_{\theta_k}\right)e^{in\phi_k}\frac{e^{\pm i
\sqrt{k_{\perp}^2-k_y^2}x'+ik_yy'+ik_zz'}}{\sqrt{k_{\perp}^2-k_y^2}},\quad x\gtrless 0.
\end{split}
\end{equation}

The  components of the translation matrices $\mathbb{U}^{12}$ and $\mathbb{U}^{21}$ are defined by
\begin{equation}\label{eq4_10_7}
\begin{split}
\mathbf{A}^{\text{TE, out}}_{nk_z}(\mathbf{x}-\mathbf{O}',k)=&\sum_{l=1}^{\infty}\sum_{m=-l}^{l}\left(U_{lm,nk_z}^{12,\text{TE,TE}}
\mathbf{A}^{\text{TE, reg}}_{lm}(\mathbf{x},k)+U_{lm,nk_z}^{12,\text{TM,TE}} \mathbf{A}^{\text{TM, reg}}_{lm}(\mathbf{x},k)\right),\\
\mathbf{A}^{\text{TM, out}}_{nk_z}(\mathbf{x}-\mathbf{O}',k)=&\sum_{l=1}^{\infty}\sum_{m=-l}^{l}\left(U_{lm,nk_z}^{12,\text{TE,TM}}
\mathbf{A}^{\text{TE, reg}}_{lm}(\mathbf{x},k)+U_{lm,nk_z}^{12,\text{TM,TM}} \mathbf{A}^{\text{TM, reg}}_{lm}(\mathbf{x},k)\right);
\end{split}
\end{equation}
\begin{equation}\label{eq4_10_9}
\begin{split}
\mathbf{A}^{\text{TE, out}}_{lm}(\mathbf{x}'+\mathbf{O}',k)=&\sum_{n=-\infty}^{\infty}H\int_{-\infty}^{\infty}\frac{dk_z}{2\pi}\left(U_{nk_z,lm}^{21,\text{TE,TE}}
\mathbf{A}^{\text{TE, reg}}_{nk_z}(\mathbf{x}',k)+U_{nk_z,lm}^{21,\text{TM,TE}} \mathbf{A}^{\text{TM, reg}}_{nk_z}(\mathbf{x}',k)\right),\\
\mathbf{A}^{\text{TM, out}}_{lm}(\mathbf{x}'+\mathbf{O}',k)=&\sum_{n=-\infty}^{\infty}H\int_{-\infty}^{\infty}\frac{dk_z}{2\pi}\left(U_{nk_z,lm}^{21,\text{TE,TM}}
\mathbf{A}^{\text{TE, reg}}_{nk_z}(\mathbf{x}',k)+U_{nk_z,lm}^{21,\text{TM,TM}} \mathbf{A}^{\text{TM, reg}}_{nk_z}(\mathbf{x}',k)\right).
\end{split}
\end{equation}
Using \eqref{eq12_4_3}, \eqref{eq12_4_4} and \eqref{eq12_4_5}, it is easy to deduce that
\begin{equation*}
U_{lm,nk_z}^{ij,\text{TE,TE}}=U_{lm,nk_z}^{ij,\text{TM,TM}},\quad U_{lm,nk_z}^{ij,\text{TE,TM}}=U_{lm,nk_z}^{ij,\text{TM,TE}},\hspace{1cm}ij=12\;\text{or}\;21.
\end{equation*}

To find the translation matrix $\mathbb{U}^{12}$, we use the fact that \cite{36}:
\begin{equation}\label{eq4_3_8}
\begin{split}
\left(\boldsymbol{\mathcal{P}}_{l''m''}\cdot\mathbf{A}_{lm}^{\text{TE,reg}}\right)(\mathbf{0})=&\frac{\mathcal{C}_{l}^{\text{reg}}}{4\pi i^{l}}
\int_0^{2\pi}d\phi_k\int_0^{\pi}
d\theta_k\,\sin\theta_k \mathbf{X}_{l'',m''}(\theta_k,\phi_k)\cdot\mathbf{X}_{lm}(\theta_k,\phi_k)=\frac{(-1)^{m+1}}{4\pi i^{l}}\delta_{l,l''}
\delta_{m'',-m}\mathcal{C}_{l}^{\text{reg}},\\
\left(\boldsymbol{\mathcal{P}}_{l''m''}\cdot\mathbf{A}_{lm}^{\text{TM,reg}}\right)(\mathbf{0})=&\frac{\mathcal{C}_{l}^{\text{reg}}}{4\pi i^{l}}
\int_0^{2\pi}d\phi_k\int_0^{\pi}
d\theta_k\,\sin\theta_k \mathbf{X}_{l'',m''}(\theta_k,\phi_k)\cdot\left(\frac{i\mathbf{k}}{k}\times\mathbf{X}_{lm}(\theta_k,\phi_k)\right)=0.
\end{split}\end{equation}
Apply  the operator $\boldsymbol{\mathcal{P}}_{l,-m}\cdot$ to both sides of \eqref{eq4_10_7} and set $\mathbf{x}=\mathbf{0}$, \eqref{eq4_3_8} implies that
\begin{equation}
\begin{split}
 &U_{lm,nk_z}^{12,\text{TE,TE}} \\=&\frac{4\pi i^{l}}{\mathcal{C}_{l}^{\text{reg}}}(-1)^{m+1}\left(\boldsymbol{\mathcal{P}}_{l,-m}\cdot
 \mathbf{A}_{nk_z}^{\text{TE,out}}\right)(-\mathbf{O}')\\
 =& \frac{2\pi(-1)^{l +1}i}{\sqrt{l(l+1)}}\sqrt{\frac{2l+1}{4\pi}\frac{(l-m)!}{(l+m)!}}\\&\times \int_{-\infty}^{\infty}dk_y \left(\frac{m}{\sin\theta_k}
 P_{l}^m(\cos\theta_k)\mathbf{e}_{\theta_k}+i\sin\theta_kP_l^{m\prime}(\cos\theta_k)\mathbf{e}_{\phi_k}\right)\cdot
 \left(-i\mathbf{e}_{\phi_k}\right)e^{i(n-m)\phi_k}\frac{e^{  i
\sqrt{k_{\perp}^2-k_y^2}L}}{\sqrt{k_{\perp}^2-k_y^2}}\\
 =& \frac{ (-1)^{l+ n+m+1}}{\sqrt{l(l+1)}}\sqrt{4\pi(2l+1)\frac{(l-m)!}{(l+m)!}}\frac{\sqrt{k^2-k_z^2}}{k}P_l^{m\prime}\left(\frac{k_z}{k}\right)\varphi_{n-m}^{\text{out}}
 (\mathbf{O}',k),\end{split}
\end{equation}
\begin{equation}
\begin{split}
& U_{lm,nk_z}^{12,\text{TE,TM}} \\=&\frac{4\pi i^{l}}{\mathcal{C}_{l}^{\text{reg}}}(-1)^{m+1}\left(\boldsymbol{\mathcal{P}}_{l,-m}\cdot
 \mathbf{A}_{nk_z}^{\text{TM,out}}\right)(-\mathbf{O}')\\
  =& \frac{2\pi(-1)^{l +1}i}{\sqrt{l(l+1)}}\sqrt{\frac{2l+1}{4\pi}\frac{(l-m)!}{(l+m)!}}\\&\times \int_{-\infty}^{\infty}dk_y \left(\frac{m}{\sin\theta_k}
 P_{l}^m(\cos\theta_k)\mathbf{e}_{\theta_k}+i\sin\theta_kP_l^{m\prime}(\cos\theta_k)\mathbf{e}_{\phi_k}\right)\cdot
 \left(-\mathbf{e}_{\theta_k}\right)e^{i(n-m)\phi_k}\frac{e^{  i
\sqrt{k_{\perp}^2-k_y^2}L}}{\sqrt{k_{\perp}^2-k_y^2}}\\
 =& \frac{ (-1)^{l+m+n } }{\sqrt{l(l+1)}}\sqrt{4\pi(2l+1)\frac{(l-m)!}{(l+m)!}}\frac{mk}{\sqrt{k^2-k_z^2}}P_l^{m}\left(\frac{k_z}{k}\right)\varphi_{n-m}^{\text{out}}
 (\mathbf{O}',k).
\end{split}
\end{equation}
Passing to imaginary frequency, we have
\begin{equation}
\begin{split}
  U_{lm,nk_z}^{12,\text{TE,TE}} (i\xi)=& \frac{ (-1)^{n }}{\sqrt{l(l+1)}}\sqrt{4\pi(2l+1)\frac{(l-m)!}{(l+m)!}}\frac{\sqrt{\kappa^2+k_z^2}}{\kappa}P_l^{m\prime}\left( \frac{ik_z}{\kappa}\right)K_{n-m}\left(L\sqrt{\kappa^2+k_z^2}\right),\\
 U_{lm,nk_z}^{12,\text{TE,TM}} (i\xi)=& \frac{ (-1)^{n } }{\sqrt{l(l+1)}}\sqrt{4\pi(2l+1)\frac{(l-m)!}{(l+m)!}}\frac{m\kappa}{\sqrt{\kappa^2+k_z^2}}P_l^{m}\left( \frac{ik_z}{\kappa}\right)K_{n-m}\left(L\sqrt{\kappa^2+k_z^2}\right).\end{split}
\end{equation}

For the translation matrix $\mathbb{U}^{12}$, we have to use the followings:
\begin{equation}\label{eq4_9_3}
\begin{split}
\left(\boldsymbol{\mathcal{Q}}_{n''}\cdot\mathbf{A}_{nk_z}^{\text{TE,reg}}\right)(\mathbf{0})=
&\frac{\mathcal{C}_n^{\text{reg}}}{2\pi i^n}\int_0^{2\pi}d\phi_k \left(-i\mathbf{e}_{\phi_k}\right)\cdot
\left(-i\mathbf{e}_{\phi_k}\right)
e^{i(n''+n)\phi_k}\\=&-\mathcal{C}_n^{\text{reg}}i^{-n}\delta_{n'',-n},\\
\left(\boldsymbol{\mathcal{Q}}_{n''}\cdot\mathbf{A}_{nk_z}^{\text{TM,reg}}\right)(\mathbf{0})=
&\frac{\mathcal{C}_n^{\text{reg}}}{2\pi i^n}\int_0^{2\pi}d\phi_k \left(-i\mathbf{e}_{\phi_k}\right)\cdot
\left( - \mathbf{e}_{\theta_k}\right)
e^{i(n''+n)\phi_k}=0.
\end{split}
\end{equation}Notice that \eqref{eq4_10_9} can be written as \begin{equation*}
\begin{split}
 &\mathcal{C}_{l}^{\text{out}}  i^{-l}\boldsymbol{\mathcal{P}}_{lm}h_0^{(1)}\left(k|\mathbf{x}'+\mathbf{O}'|\right)\\
=&\sum_{n=-\infty}^{\infty}H\int_{-\infty}^{\infty}\frac{dk_z}{2\pi}\frac{\mathcal{C}_n^{\text{reg}}}{2\pi i^n}
\int_0^{2\pi}d\phi_k \left( U_{nk_z,lm}^{21,\text{TE,TE}}
\left( -i \mathbf{e}_{\phi_k}\right)+U_{nk_z,lm}^{21,\text{TM,TE}} \left(-
 \mathbf{e}_{\theta_k}\right)\right)\\&\hspace{7cm}\times e^{in\phi_k}
e^{i\sqrt{k^2-k_z^2}\cos\phi_kx'+i\sqrt{k^2-k_z^2}\sin\phi_ky'+ik_zz'}.
\end{split}
\end{equation*}
Let us concentrate on the case where $m\geq 0$. The case where $m<0$ can be worked out in the same way.
As in the scalar case, passing to imaginary frequency and taking Fourier transform, we find that
\begin{equation}\label{eq5_3_5}
\begin{split}
&\sum_{n=-\infty}^{\infty} \frac{1}{2\pi  }
\int_0^{2\pi}d\phi_k \left( U_{nk_z,lm}^{21,\text{TE,TE}}(i\xi)
\left( -i \mathbf{e}_{\phi_k}\right)+U_{nk_z,lm}^{21,\text{TM,TE}}(i\xi)\left(-
 \mathbf{e}_{\theta_k}\right)\right)\\&\hspace{7cm}\times e^{in\phi_k}
e^{-\sqrt{\kappa^2+k_z^2} \cos\phi_k x'-\sqrt{\kappa^2+k_z^2}\sin\phi_k y'  }\\
=&\frac{\pi (-1)^n}{2H\kappa}\sqrt{\frac{2l+1}{4\pi l(l+1)}\frac{(l-m)!}{(l+m)!}}\Biggl(\frac{\mathbf{e}_x}{2}
\left[P_l^{m+1}\left(-\frac{ik_z}{\kappa}\right)\mathcal{Q}_{m+1}+(l+m)(l-m+1)P_l^{m-1}\left(-\frac{ik_z}{\kappa}\right)\mathcal{Q}_{m-1}\right]
\\
&\hspace{4cm}+\frac{\mathbf{e}_y}{2i}\left[P_l^{m+1}\left(-\frac{ik_z}{\kappa}\right)\mathcal{Q}_{m+1}-(l+m)(l-m+1)P_l^{m-1}\left(-\frac{ik_z}{\kappa}\right)
\mathcal{Q}_{m-1}\right]\\
&\hspace{4cm}+m\mathbf{e}_zP_l^{m}\left(-\frac{ik_z}{\kappa}\right)\mathcal{Q}_{m}\Biggr)\int_{-\infty}^{\infty}dk_y\frac{e^{- \sqrt{\kappa^2+k_z^2+k_y^2}(x'+L)+ik_yy'}}{\sqrt{\kappa^2+k_z^2+k_y^2}}.
\end{split}\end{equation}
Here we have used the formula \eqref{eq5_3_4}. Apply the operator $\boldsymbol{\mathcal{Q}}_{-n}\cdot$ to both sides of \eqref{eq5_3_5} and set $x'=y'=0$, \eqref{eq4_9_3} gives
\begin{equation}\label{eq5_3_6}
\begin{split}
U_{nk_z,lm}^{21,\text{TE,TE}}(i\xi)
=&\frac{\pi (-1)^{n+1}}{4H\kappa}\sqrt{\frac{2l+1}{4\pi l(l+1)}\frac{(l-m)!}{(l+m)!}}\Biggl(-P_l^{m+1}\left(-\frac{ik_z}{\kappa}\right)+(l+m)(l-m+1)P_l^{m-1}\left(-\frac{ik_z}{\kappa}\right)\Biggr)\\
& \times\int_{-\infty}^{\infty}dk_ye^{i(m-n)\phi_k}\frac{e^{- \sqrt{\kappa^2+k_z^2+k_y^2}L}}{\sqrt{\kappa^2+k_z^2+k_y^2}}\\
=&\frac{\pi(-1)^{n+1 }}{ H\kappa}\frac{\sqrt{\kappa^2+k_z^2}}{\kappa}\sqrt{\frac{2l+1}{4\pi l(l+1)}\frac{(l-m)!}{(l+m)!}} P_l^{m\prime}\left(-\frac{ik_z}{\kappa}\right)\varphi_{m-n,k_z}^{\text{out}}(\mathbf{O}',i\kappa).
\end{split}\end{equation}
Applying
$$\left[\frac{\mathbf{e}_x}{2}\left(\mathcal{Q}_{-n+1}+\mathcal{Q}_{-n-1}\right)+\frac{\mathbf{e}_y}{2i}\left(\mathcal{Q}_{-n+1}-\mathcal{Q}_{-n-1}\right)\right]\cdot$$ to  both sides of \eqref{eq5_3_5} and set $x'=y'=0$, one find that
\begin{equation}\label{eq5_3_7}
\begin{split}
U_{nk_z,lm}^{21,\text{TM,TE}}(i\xi)
=&\frac{\pi (-1)^{n+1}i}{4H\kappa}\sqrt{\frac{2l+1}{4\pi l(l+1)}\frac{(l-m)!}{(l+m)!}}\frac{\kappa}{k_z}\Biggl(P_l^{m+1}\left(-\frac{ik_z}{\kappa}\right)+(l+m)(l-m+1)P_l^{m-1}\left(-\frac{ik_z}{\kappa}\right)\Biggr)\\
&\times \int_{-\infty}^{\infty}dk_ye^{i(m-n)\phi_k}\frac{e^{- \sqrt{\kappa^2+k_z^2+k_y^2}L}}{\sqrt{\kappa^2+k_z^2+k_y^2}}\\
=&\frac{\pi(-1)^{n +1}  }{ H\kappa}\frac{m\kappa}{\sqrt{\kappa^2+k_z^2}}\sqrt{\frac{2l+1}{4\pi l(l+1)}\frac{(l-m)!}{(l+m)!}} P_l^{m}\left(-\frac{ik_z}{\kappa}\right)\varphi_{m-n,k_z}^{\text{out}}(\mathbf{O}',i\kappa).
\end{split}\end{equation}
Hence,
\begin{equation}
\begin{split}
U_{nk_z,lm}^{21,\text{TE,TE}}
(L\mathbf{e}_x,i\xi) = & \frac{(-1)^{n +1}\pi  }{H\kappa}\sqrt{\frac{2l+1}{4\pi l(l+1)}\frac{(l-m)!}{(l+m)!}}
\frac{\sqrt{\kappa^2+k_z^2}}{\kappa}P_l^m\left(-\frac{ik_z}{\kappa}\right)K_{m-n}\left(L\sqrt{\kappa^2+k_z^2} \right),\\
U_{nk_z,lm}^{21,\text{TM,TE}}
(L\mathbf{e}_x,i\xi) = & \frac{(-1)^{n +1}\pi   }{H\kappa}\sqrt{\frac{2l+1}{4\pi l(l+1)}\frac{(l-m)!}{(l+m)!}}
\frac{m\kappa}{\sqrt{\kappa^2+k_z^2}}P_l^m\left(-\frac{ik_z}{\kappa}\right)K_{m-n}\left(L\sqrt{\kappa^2+k_z^2} \right).
\end{split}\end{equation}
Finally, we can write down the formula for  the electromagnetic Casimir interaction energy between a perfectly conducting sphere and a perfectly conducting cylinder. It is given by \eqref{eq3_20_7} with \begin{equation*}
\begin{split}
\mathbb{M}_{lm,l'm'}(i\xi)=&-\frac{\mathbb{T}^{lm}}{2 }\sqrt{ \frac{(2l+1)(2l'+1)}{l(l+1)l'(l'+1)}\frac{(l-m)!}{(l+m)!}\frac{(l'-m')!}{(l'+m')!}}\\
&\times\sum_{n=-\infty}^{\infty}
\int_{-\infty}^{\infty} d \theta \cosh\theta
\left(\begin{aligned} \cosh\theta P_l^{m\prime}(i\sinh\theta)\hspace{0.3cm} &\hspace{0.3cm}
\frac{m}{\cosh\theta}P_l^m(i\sinh\theta)\\
\frac{m}{\cosh\theta}P_l^m(i\sinh\theta)\hspace{0.3cm} &\hspace{0.5cm}\cosh\theta P_l^{m\prime}(i\sinh\theta)\end{aligned}\right)
\mathbb{T}^{nk_z} \\&\times \left(\begin{aligned} \cosh\theta P_{l'}^{m'\prime}(-i\sinh\theta)\hspace{0.3cm} &\hspace{0.3cm}
\frac{m'}{\cosh\theta}P_{l'}^{m'}(-i\sinh\theta)\\
\frac{m'}{\cosh\theta}P_{l'}^{m'}(-i\sinh\theta)\hspace{0.3cm} &\hspace{0.5cm}\cosh\theta P_{l'}^{m'\prime}(-i\sinh\theta)\end{aligned}\right)K_{n-m}\left(\kappa L\cosh\theta\right)
K_{m'-n}\left(\kappa L\cosh\theta\right).
\end{split}\end{equation*}
The trace Tr in \eqref{eq3_20_7} is now given by
\begin{equation*}
\text{Tr}=\sum_{l=1}^{\infty}\sum_{m=-l}^{l} \text{tr},
\end{equation*}
where tr is the trace over $2\times 2$ matrices.

As in the previous section, one can show that the Casimir interaction energy is real. However, it is not straightforward that it is always negative.
   \section{Large separation   behavior }\label{sec4}
In this section, we compute the leading term of the Casimir energy when the separation between the sphere and the cylinder is much larger than their respective radii, namely, when $$L\gg R_1, R_2.$$
We need to use the following leading behaviors for the modified Bessel functions: When $z\rightarrow 0$,
 \begin{equation}\label{eq12_6_3}
\begin{split}
&\frac{I_0(z)}{K_0(z)}\sim  -\frac{1}{\ln z}+\ldots,\\
&\frac{I_0'(z)}{K_0'(z)}\sim  -\frac{z^2}{2}+\ldots,\\
&\frac{I_1'(z)}{K_1'(z)}\sim  -\frac{z^2}{2}+\ldots,\\
&\frac{I_{\frac{1}{2}}(z)}{K_{\frac{1}{2}}(z)}\sim  \frac{2}{\pi  }z+\ldots,\\
&\frac{I_{\frac{3}{2}}(z)}{K_{\frac{3}{2}}(z)}\sim \frac{2}{3\pi}z^3+\ldots,\\
&\frac{-\frac{1}{2}I_{\frac{1}{2}}(z)+zI_{\frac{1}{2}}'(z)}{-\frac{1}{2}K_{\frac{1}{2}}(z)+zK_{\frac{1}{2}}'(z)}\sim -\frac{2}{3\pi}z^3+\ldots,\\
&\frac{-\frac{1}{2}I_{\frac{3}{2}}(z)+zI_{\frac{3}{2}}'(z)}{-\frac{1}{2}K_{\frac{3}{2}}(z)+zK_{\frac{3}{2}}'(z)}\sim -\frac{1}{3\pi}z^3+\ldots,\\
&\frac{\frac{1}{2}I_{\frac{3}{2}}(z)+zI_{\frac{3}{2}}'(z)}{\frac{1}{2}K_{\frac{3}{2}}(z)+zK_{\frac{3}{2}}'(z)}\sim -\frac{4}{3\pi}z^3+\ldots.
\end{split}
\end{equation}

 Making a change of variables
$$\xi\mapsto \omega=\frac{L\xi}{c}$$in \eqref{eq12_5_6}, we have
\begin{equation}\label{eq12_7_1}\begin{split}
E_{\text{Cas}}
=&-\sum_{s=1}^{\infty}\frac{1}{s}\frac{\hbar c}{2\pi L}\int_0^{\infty} d\omega \left(\sum_{l_i=0}^{\infty}\sum_{m_i=-l_i}^{l_i} \right)\prod_{i=1}^sM_{l_im_i,l_{i+1}m_{i+1}}\left(\frac{ic\omega}{L}\right).
\end{split}
\end{equation}

The large separation leading term comes from the term which has the lowest power of $\omega$. For the case Dirichlet boundary conditions are imposed on both the sphere and the cylinder, the large separation leading term comes from the term with $s=1$, $l_1=m_1=0$. Namely,
\begin{equation*}
\begin{split}
E_{\text{Cas}}^{\text{D}}\approx& -\frac{\hbar c}{2\pi L}\int_0^{\infty}d\omega
M_{00,00}^{\text{D}}.
\end{split}
\end{equation*}Moreover, the leading term of $M_{00,00}^{\text{D}}$ comes from the term with $n=0$, i.e.,
\begin{equation*}
\begin{split}
M_{00,00}^{\text{D}}\approx &\frac{1}{2 }\frac{I_{ \frac{1}{2}}(a\omega)}{K_{ \frac{1}{2}}(a\omega)}
\int_{-\infty}^{\infty} d \theta \cosh\theta
K_{0}\left( \omega\cosh\theta\right)\frac{I_{0}(b\omega\cosh\theta)}{K_{0}(b\omega\cosh\theta)}
 K_{0}\left( \omega\cosh\theta\right).
\end{split}\end{equation*}Here
$$a=\frac{R_1}{L},\quad b=\frac{R_2}{L}.$$ When $R_1,R_2\ll L$, we obtain from \eqref{eq12_6_3} that
\begin{equation*}\begin{split}
M_{00,00}^{\text{D}}\approx &-\frac{a\omega}{\pi\ln b}\int_{-\infty}^{\infty} d \theta \cosh\theta
K_{0}\left( \omega\cosh\theta\right)
 K_{0}\left( \omega\cosh\theta\right).
\end{split}
\end{equation*}
Therefore, the leading term of the Casimir interaction energy is
\begin{equation*}
\begin{split}
E_{\text{Cas}}^{\text{D}}\approx&-\frac{\hbar c R_1}{ 2\pi^2 L^2 \ln (L/R_2)}\int_0^{\infty}d\omega
 \omega
\int_{-\infty}^{\infty} d \theta \cosh\theta
K_{0}\left( \omega\cosh\theta\right)
 K_{0}\left( \omega\cosh\theta\right)\\
 =
 &-\frac{ \hbar c R_1}{ 4\pi^2 L^2 \ln (L/R_2)}\int_{-\infty}^{\infty}d\theta\frac{1}{\cosh\theta}\\
 =
 &-\frac{ \hbar c R_1}{ 4\pi L^2\ln (L/R_2)}.
\end{split}
\end{equation*}

For the case Neumann boundary conditions are imposed on both the sphere and the cylinder, the large distance leading term comes from the term with $s=1$, $l_1=m_1=0$ or $l_1=1$, $m_1=0,\pm 1$. Namely,
\begin{equation*}
\begin{split}
E_{\text{Cas}}^{\text{N}}\approx& -\frac{\hbar c}{2\pi L}\int_0^{\infty}d\omega
\left(M_{00,00}^{\text{N}}+M_{10,10}^{\text{N}}+M_{11,11}^{\text{N}}+M_{1,-1;1,-1}^{\text{N}}\right).
\end{split}
\end{equation*}Now for each of these $M_{lm,lm}^{\text{N}}$, the leading term comes from $n=0,\pm 1$.
Using \eqref{eq12_6_3}, we have
\begin{equation*}
\begin{split}
\int_0^{\infty}d\omega M_{00,00}^{\text{N}}\approx  &\frac{R_1^3R_2^2}{6\pi L^5}\int_0^{\infty}d\omega \omega^5\int_{-\infty}^{\infty} d \theta \cosh^3\theta\left( K_{0}\left( \omega\cosh\theta\right)^2+2 K_{1}\left( \omega\cosh\theta\right)^2\right)\\
=&\frac{16R_1^3R_2^2}{45 L^5},
\\
\int_0^{\infty}d\omega  M_{10,10}^{\text{N}}\approx  &\frac{R_1^3R_2^2}{4\pi L^5}\int_0^{\infty}d\omega \omega^5\int_{-\infty}^{\infty} d \theta \cosh^3\theta\sinh^2\theta\left( K_{0}\left( \omega\cosh\theta\right)^2+2 K_{1}\left( \omega\cosh\theta\right)^2\right)\\
=&\frac{8 R_1^3R_2^2}{15 L^5},
 \\
\int_0^{\infty}d\omega M_{1\pm1,1\pm1}^{\text{N}}
 \approx &\frac{R_1^3R_2^2}{8\pi L^5}\int_0^{\infty}d\omega\omega^5\int_{-\infty}^{\infty} d \theta \cosh^5\theta \left( K_{0}\left( \omega\cosh\theta\right)^2+  K_{1}\left( \omega\cosh\theta\right)^2+  K_{2}\left( \omega\cosh\theta\right)^2\right)\\
 =&\frac{17 R_1^3R_2^2}{15 L^5}
 \end{split}\end{equation*}
 Hence, the leading term of the Casimir interaction energy is
 \begin{equation*}
\begin{split}
E_{\text{Cas}}^{\text{N}}\approx& -\frac{\hbar c}{2\pi L}\left(\frac{16R_1^3R_2^2}{45 L^5}+\frac{8 R_1^3R_2^2}{15 L^5}+\frac{34 R_1^3R_2^2}{15L^5}\right)=-\frac{71 \hbar c R_1^3R_2^2 }{45\pi L^6}.
\end{split}
\end{equation*}

For the case perfectly conducting boundary conditions are imposed on both the sphere and the cylinder, the large distance leading term comes from the term with $s=1$, $l_1=1$ and $m_1=0,\pm 1$. Namely,
\begin{equation*}
\begin{split}
E_{\text{Cas}}^{\text{P}}\approx& -\frac{\hbar c}{2\pi L}\int_0^{\infty}d\omega \,\text{tr}
 \left(\mathbb{M}_{10,10} +\mathbb{M}_{11,11} +\mathbb{M}_{1,-1;1,-1} \right).
\end{split}
\end{equation*}For each of these $\mathbb{M}_{lm,lm}$, the leading term comes from $n=0$.
\begin{equation*}
\begin{split}
\int_0^{\infty}d\omega \,\text{tr}\,\mathbb{M}_{10,10} \approx &\frac{1}{ \pi}\frac{R_1^3}{L^3\ln (L/R_2)}
\int_0^{\infty}d\omega \omega^3\int_{-\infty}^{\infty} d \theta \cosh^3\theta
   K_{0}\left( \omega\cosh\theta\right)^2\\=&\frac{R_1^3}{3L^3\ln (L/R_2)}, \\
\int_0^{\infty}d\omega \,\text{tr}\,\mathbb{M}_{1\pm 1,1\pm 1} \approx & \frac{1}{ 4\pi}\frac{R_1^3}{L^3\ln (L/R_2)}\int_0^{\infty}d\omega \omega^3
\int_{-\infty}^{\infty} d \theta \cosh\theta(2\sinh^2\theta-1)
   K_{1}\left( \omega\cosh\theta\right)^2\\
   =&\frac{R_1^3}{12L^3\ln (L/R_2)}. \end{split}\end{equation*}
  Therefore, the leading term of the Casimir interaction energy is
 \begin{equation*}
\begin{split}
E_{\text{Cas}}^{\text{P}}\approx& -\frac{\hbar c }{2\pi L}\left(\frac{R_1^3}{3L^3\ln (L/R_2)}+\frac{R_1^3}{6L^3\ln (L/R_2)}\right)\\
=&-\frac{\hbar cR_1^3}{4\pi L^4\ln (L/R_2)}.
\end{split}
\end{equation*}

In summary, the large separation leading terms for Dirichlet, Neumann and perfectly boundary conditions are given respectively by
\begin{equation*}
\begin{split}
E_{\text{Cas}}^{\infty,\text{D}}=& -\frac{\hbar cR_1}{ 4\pi L^2\ln (L/R_2)},\\
E_{\text{Cas}}^{\infty,\text{N}}=& -\frac{71 \hbar cR_1^3R_2^2}{45\pi L^6},\\
E_{\text{Cas}}^{\infty,\text{P}}=& -\frac{\hbar c R_1^3}{4\pi L^4\ln(L/R_2)}.
\end{split}
\end{equation*}
It is interesting to compare these to the large separation leading terms between two spheres and between two cylinders. For two spheres with radii $R_1$ and $R_2$, the large separation leading terms for Dirichlet, Neumann and perfectly conducting boundary conditions are given respectively by \cite{15,21,33}:
\begin{equation*}
\begin{split}
E_{\text{Cas}}^{\infty,\text{D}}=&  -\frac{\hbar cR_1R_2}{4\pi L^3 },\\
E_{\text{Cas}}^{\infty,\text{N}}=& -\frac{ 161\hbar cR_1^3R_2^3}{96\pi L^7},\\
E_{\text{Cas}}^{\infty,\text{P}}=& -\frac{ 143\hbar cR_1^3R_2^3}{16\pi L^7}.
\end{split}
\end{equation*}
For two cylinders with radii $R_1$ and $R_2$ and length $H$, the large separation leading terms for Dirichlet, Neumann and perfectly conducting boundary conditions are given respectively by \cite{19}:
\begin{equation*}
\begin{split}
E_{\text{Cas}}^{\infty,\text{D}}=&  -\frac{\hbar cH}{8\pi L^2\left[\ln (L/R_1)\right]\left[\ln (L/R_2)\right]},\\
E_{\text{Cas}}^{\infty,\text{N}}=& -\frac{7\hbar cHR_1^2R_2^2}{5\pi L^6},\\
E_{\text{Cas}}^{\infty,\text{P}}=& -\frac{\hbar cH}{8\pi L^2\left[\ln (L/R_1)\right]\left[\ln (L/R_2)\right]}.
\end{split}
\end{equation*}For any boundary conditions, we find that the leading term of the Casimir interaction energy for the sphere-cylinder configuration is intermediate between the sphere-sphere configuration and the cylinder-cylinder configuration.

\begin{figure}[h]
\epsfxsize=0.49\linewidth \epsffile{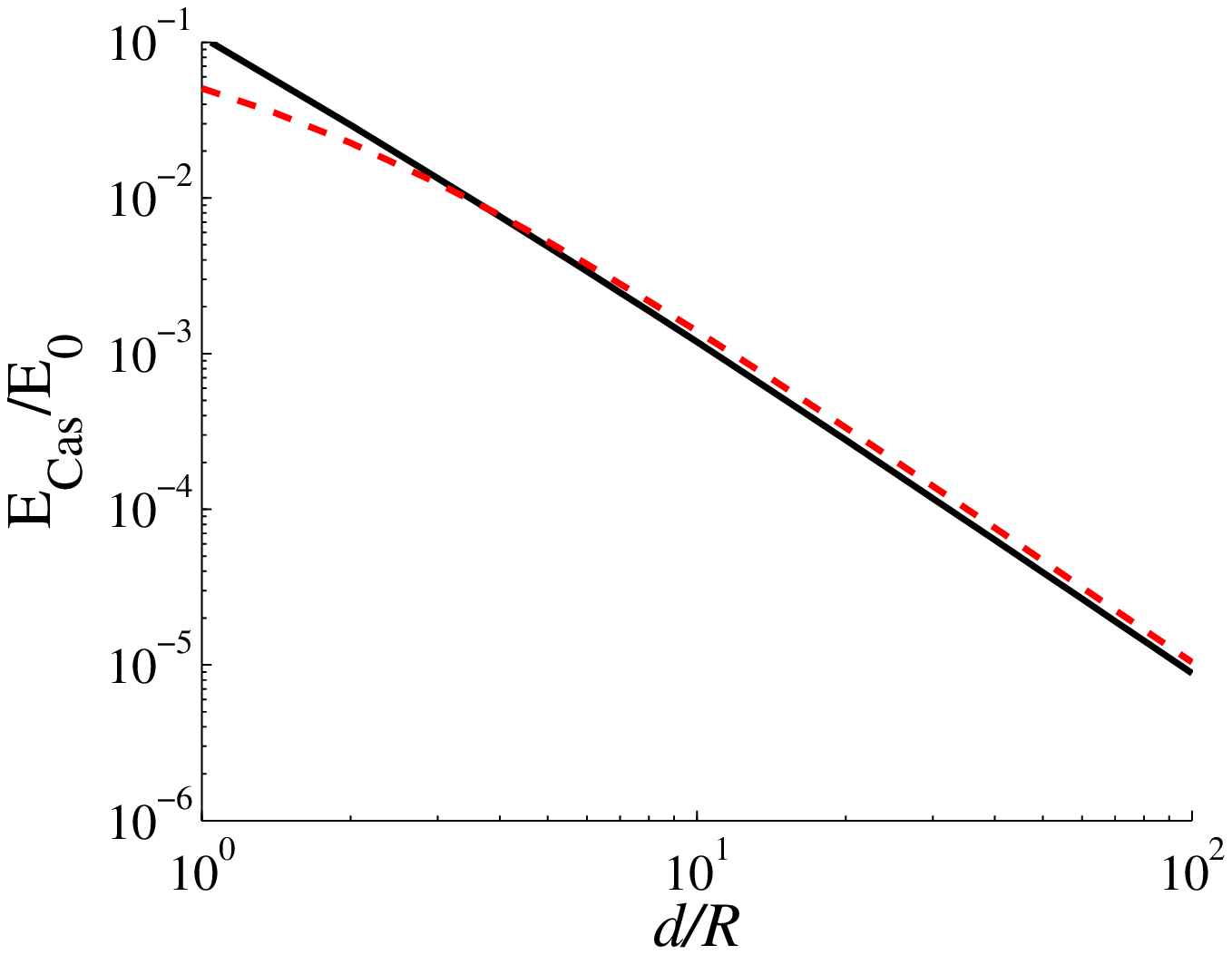}\epsfxsize=0.49\linewidth \epsffile{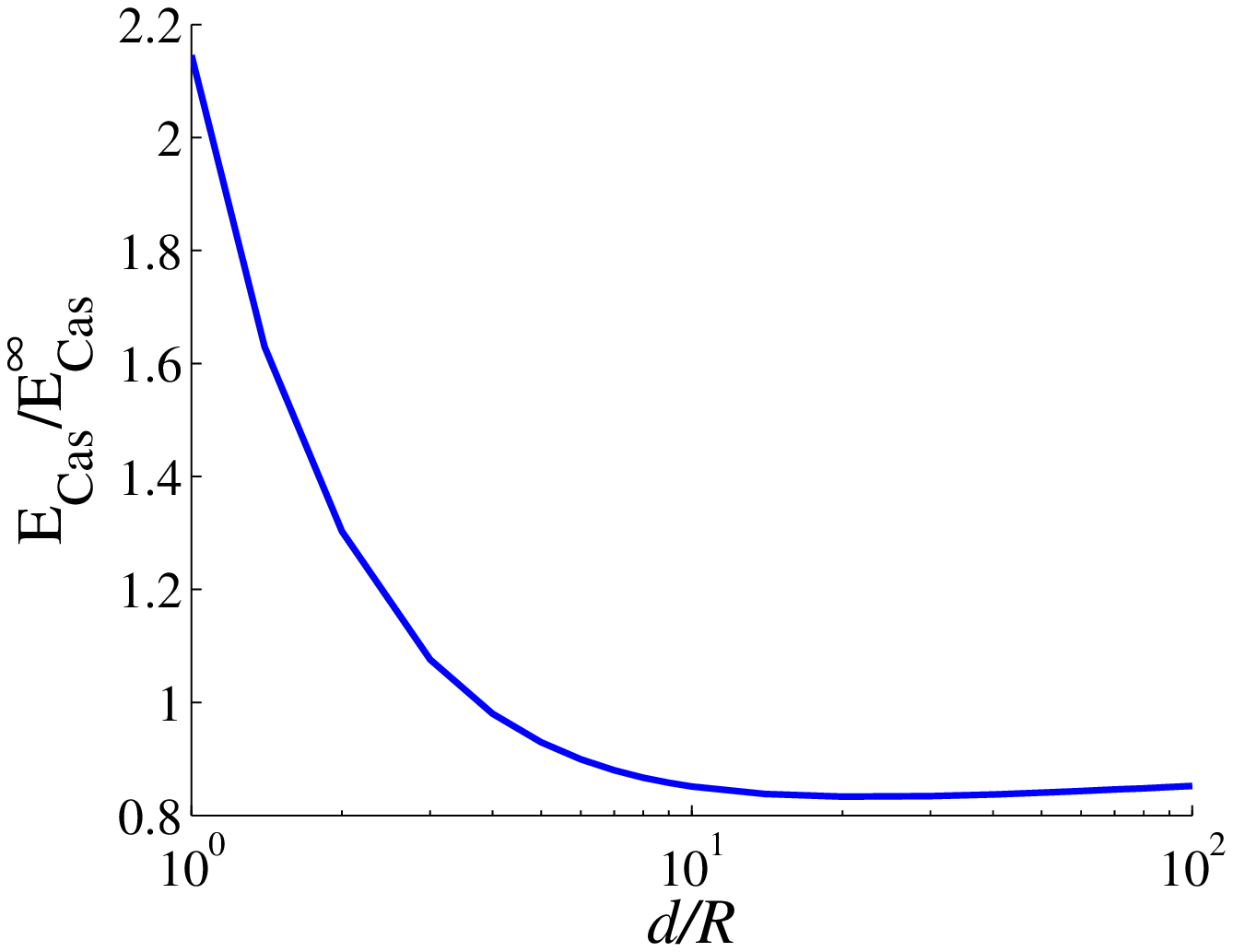}  \caption{\label{f2} In the figure on the left, the solid line shows the exact Dirichlet Casimir interaction energy $E_{\text{Cas}}^{\text{D}}$ normalized by $E_0=\hbar c/(2\pi R)$ as a function of $d/R$. The dotted line shows the large separation leading term $E_{\text{Cas}}^{\infty,\text{D}}$ normalized by $E_0$. The figure on the right shows the ratio of the exact Casimir interaction energy to the large separation leading term.}\end{figure}

\begin{figure}[h]
\epsfxsize=0.49\linewidth \epsffile{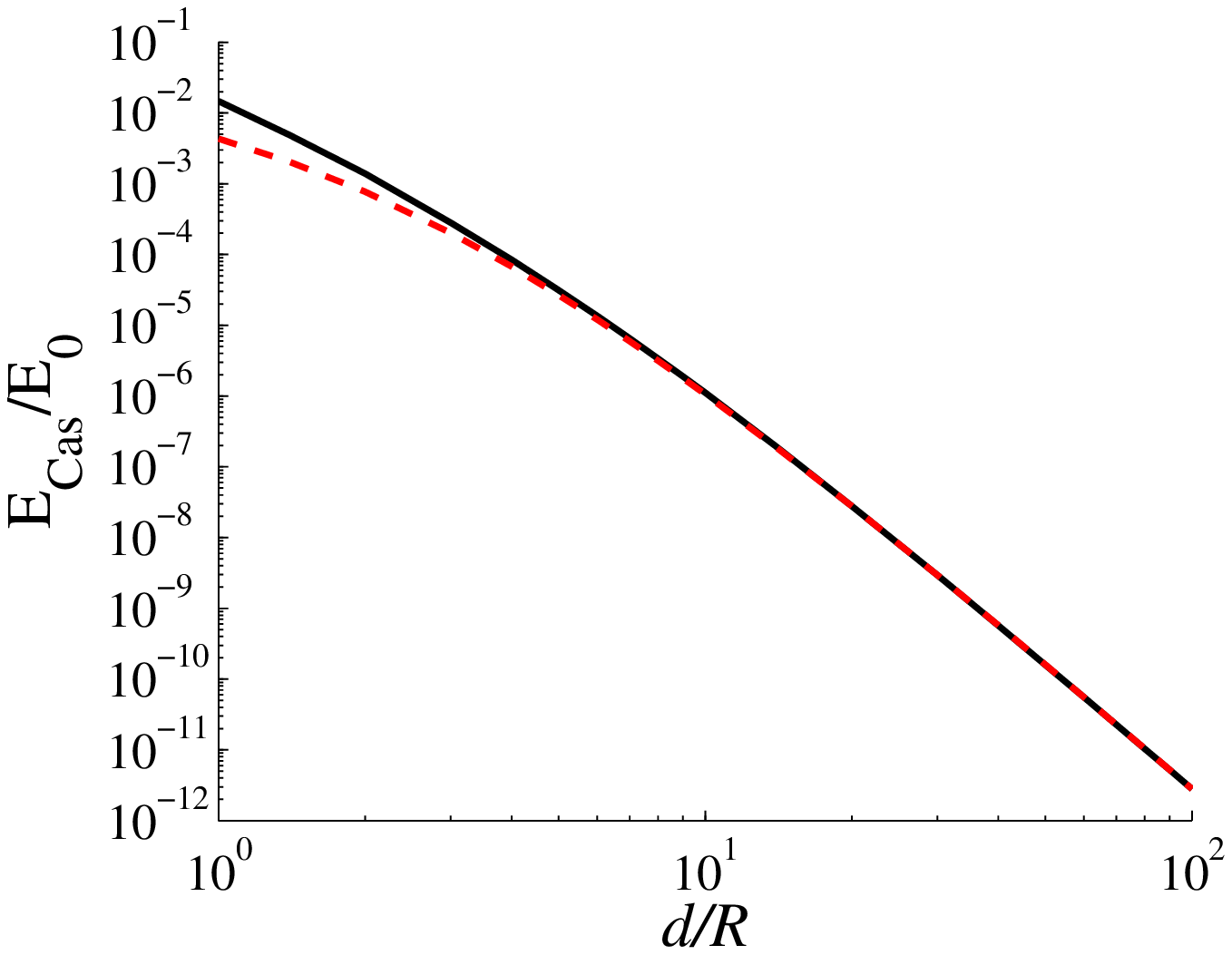}\epsfxsize=0.49\linewidth \epsffile{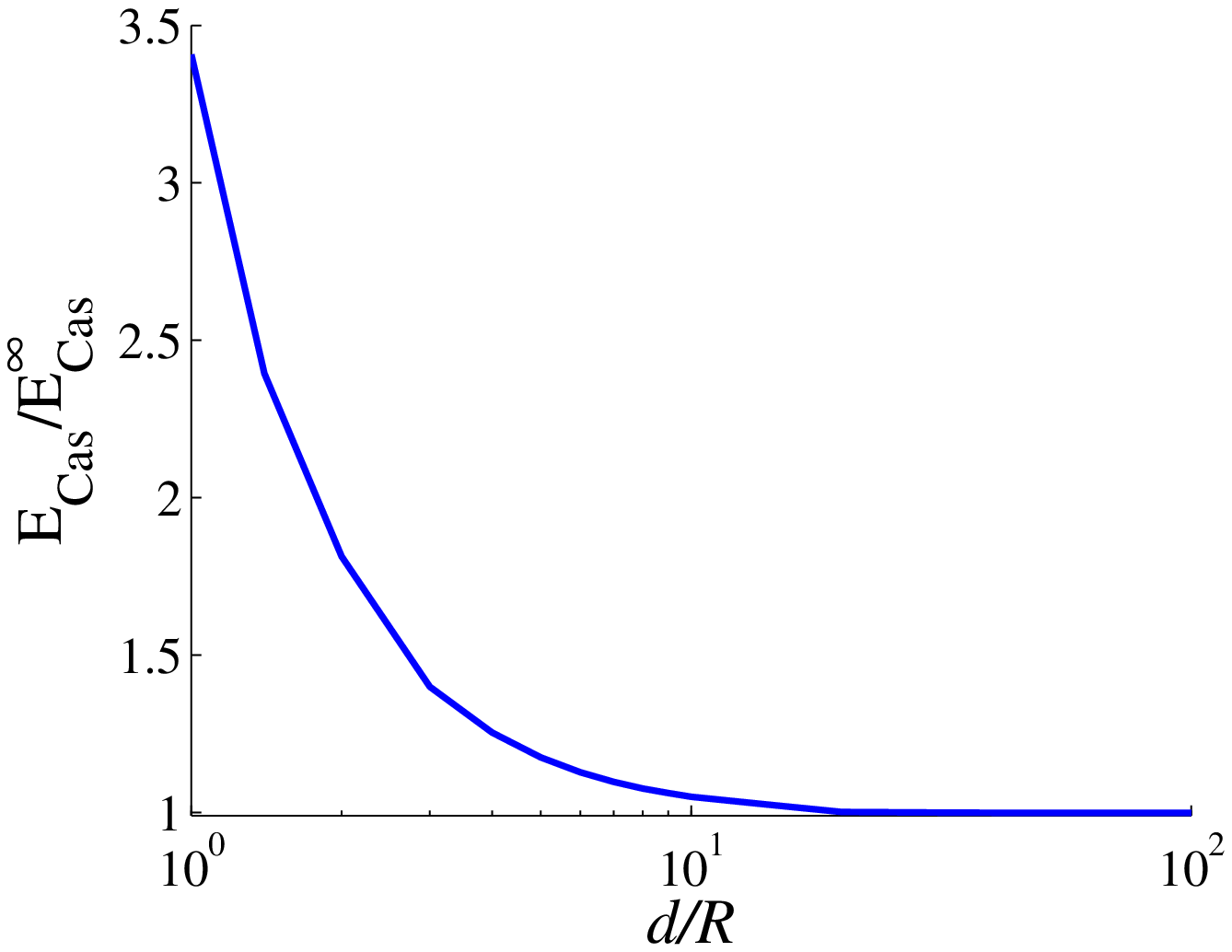}  \caption{\label{f3} In the figure on the left, the solid line shows the exact Neumann Casimir interaction energy $E_{\text{Cas}}^{\text{N}}$ normalized by $E_0=\hbar c/(2\pi R)$ as a function of $d/R$. The dotted line shows the large separation leading term $E_{\text{Cas}}^{\infty,\text{N}}$ normalized by $E_0$. The figure on the right shows the ratio of the exact Casimir interaction energy to the large separation leading term.}\end{figure}

\begin{figure}[h]
\epsfxsize=0.49\linewidth \epsffile{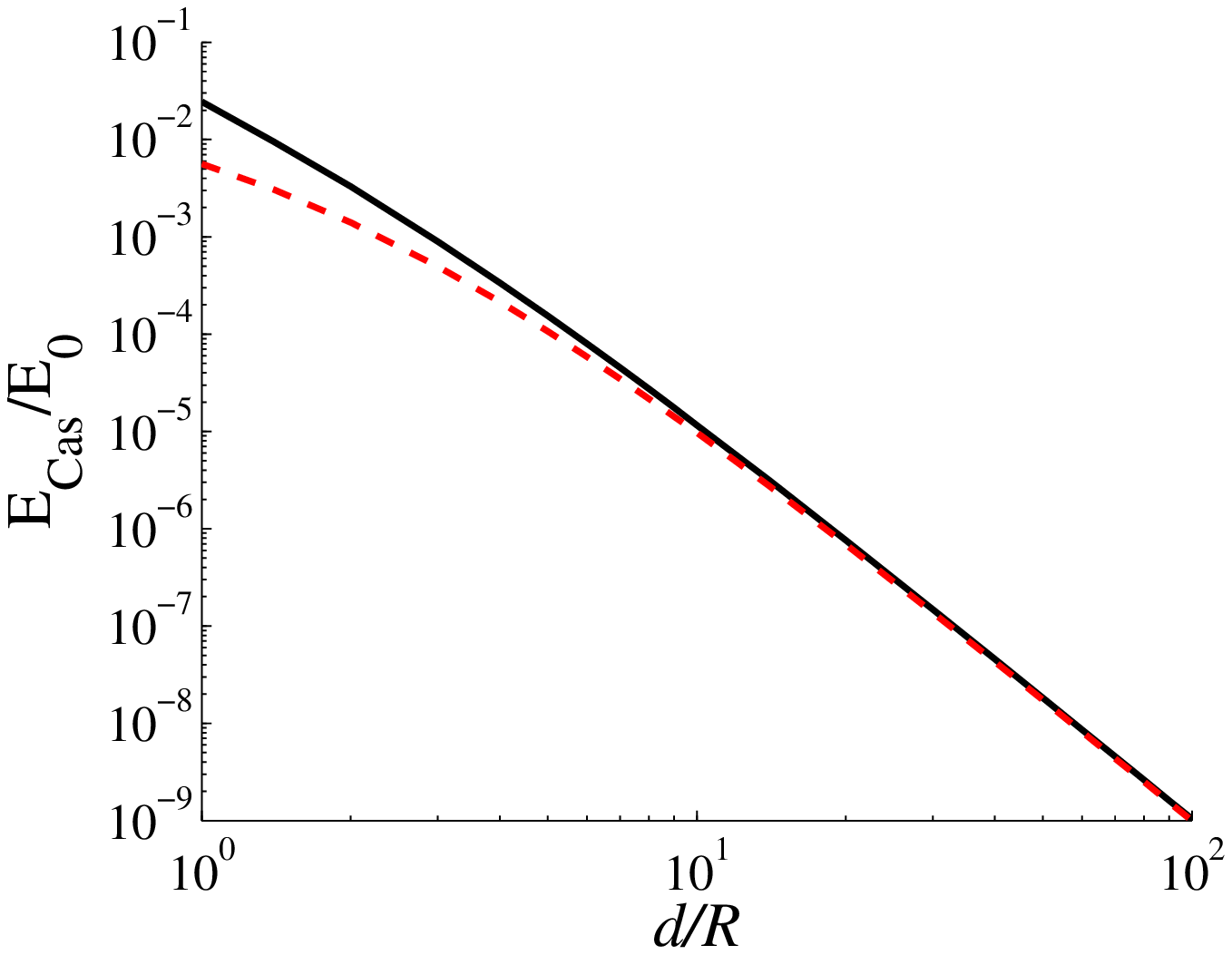}\epsfxsize=0.49\linewidth \epsffile{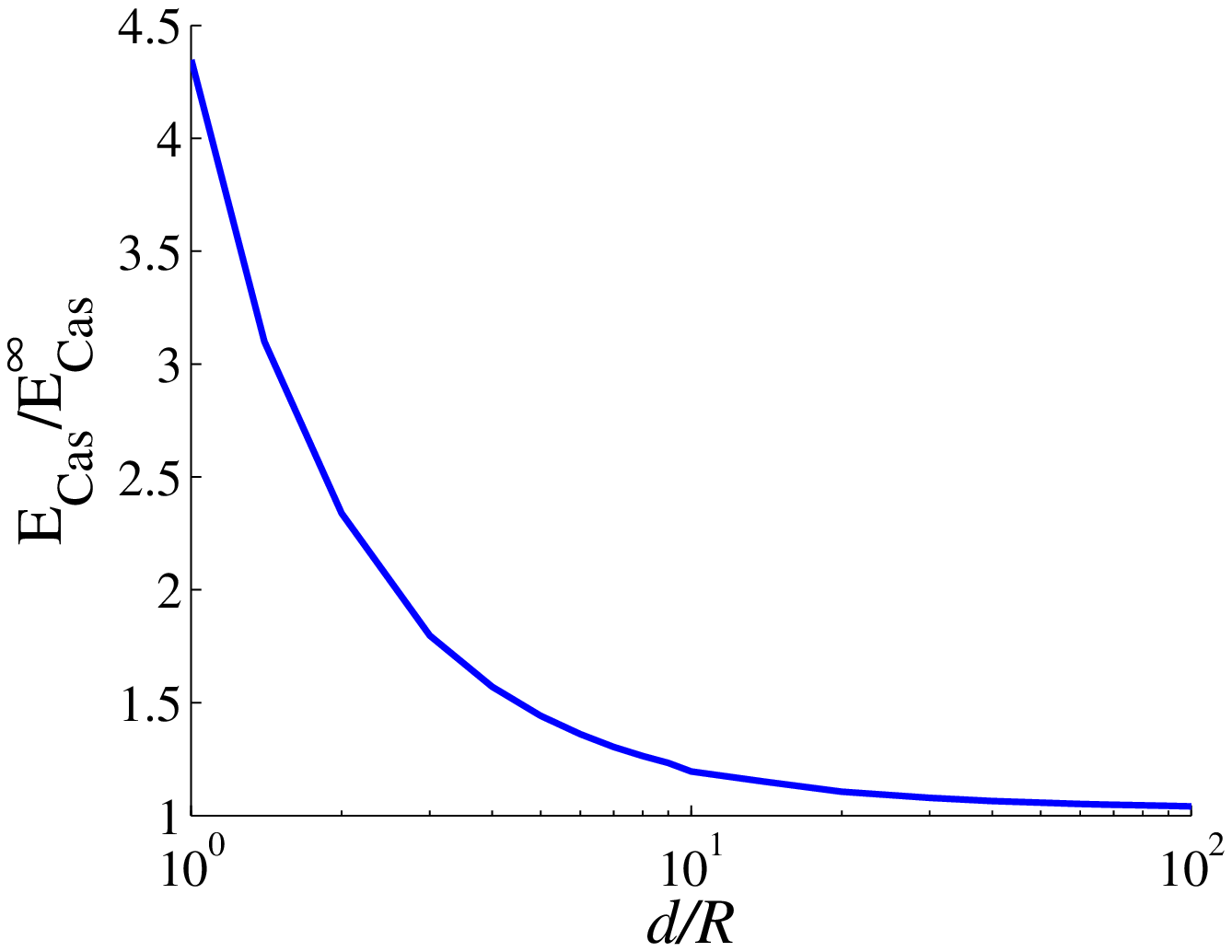}  \caption{\label{f4} In the figure on the left, the solid line shows the  exact Casimir interaction energy $E_{\text{Cas}}^{\text{P}}$ for perfectly conducting boundary conditions normalized by $E_0=\hbar c/(2\pi R)$ as a function of $d/R$. The dotted line shows the large separation leading term $E_{\text{Cas}}^{\infty,\text{P}}$ normalized by $E_0$. The figure on the right shows the ratio of the exact Casimir interaction energy to the large separation leading term.}\end{figure}

Numerical evaluations of the Casimir interaction energy $E_{\text{Cas}}$   for Dirichlet, Neumann and perfectly conducting boundary conditions are plotted respectively in Fig. \ref{f2}, Fig. \ref{f3} and Fig. \ref{f4}, and they are compared to the respective large separation leading term $E^{\infty}_{\text{Cas}}$ derived above. The  energies are normalized by $E_0=\hbar c/(2\pi R)$ and they are plotted as   functions of $d/R$ for $1\leq d/R\leq 100$, where $d$ is the distance between the sphere and the cylinder, and $R$ is their common radius. In all the cases, we see that the large separation leading term agrees quite well with the exact value when $d/R$ is large.

\section{Conclusion}\label{sec6}
In this work, we discuss the Casimir interaction between a sphere and a cylinder, which has not been considered before. The sphere-cylinder configuration can play the same role as the sphere-plane configuration in Casimir experiments, since it does not have the problem of maintaining parallelism in the plane-plane configuration.
We consider scalar interaction with Dirichlet or Neumann boundary conditions, and electromagnetic interaction with perfectly conducting boundary conditions.

We first derive the exact formula for the Casimir interaction energy. The hardest part in the derivation is the computation of the translation matrices between the sphere and the cylinder. Motivated by the operator method used by Wittman \cite{36} to derive the translation matrices between two spheres, we have derived the translation matrices between a sphere and a plane and a cylinder and a plane in \cite{37}. In this work, we use the same method to derive the translation matrices between a sphere and a cylinder. The results can be expressed in terms of modified Bessel functions and associated Legendre functions. The final exact formula for the Casimir interaction energy is quite complicated. Nevertheless,    when both the sphere and the cylinder are imposed with Dirichlet boundary conditions or Neumann boundary conditions, we can deduce from the exact formula that the Casimir force is always attractive.

Using the exact formula, we compute the leading terms of the Casimir interaction energy when the separation between the sphere and the cylinder is large. It behaves like $\sim \hbar c R_1/[L^2\ln(L/R_2)]$, $\sim \hbar c R_1^3R_2^2/L^6$ and $\sim \hbar c R_1^3/[L^4\ln(L/R_2)]$ respectively for Dirichlet, Neumann and perfectly conducting boundary conditions. Here $R_1$ and $R_2$ are   the radii of the   sphere and the cylinder respectively, and $L$ is the distance between the centers of the sphere and the cylinder. As for other configurations, we find that at large separation,  the Casimir force is strongest in the Dirichlet case, and is weakest in the Neumann case.

 Since Casimir force has been confirmed in the sphere-plane configuration,    measuring Casimir force between a sphere and a cylinder should be possible. We  believe that this configuration will play an important role in nanotechnology.
\begin{acknowledgments}\noindent
 This work is supported by the Ministry of Higher Education of Malaysia  under the FRGS grant FRGS/2/2010/SG/UNIM/02/2. We would like to thank the anonymous referee for the helpful comments to improve this article.

\end{acknowledgments}

\appendix
 \section{Small separation behavior}\label{sec5}

In this section, we consider the small distance behavior of the Casimir interaction energy. In principle, one can compute the small distance asymptotic expansion of the Casimir interaction energy from the exact formula using the perturbative method developed in \cite{25}. However, this approach is very complicated and is beyond the scope of this work. Here we compute the leading order term from the proximity force approximation and the next to leading order term using the derivative expansion proposed in \cite{43}.

As is well-known, in the small distance limit, the leading term of the Casimir interaction energy is given by the proximity force approximation \cite{39,40}.
Based on the cylinder, which we parametrize by $x=R_2\cos\phi+L$, $y=R_2\sin\phi$ and $z=z$, where $0\leq \phi\leq 2\pi$ and $-H/2\leq z\leq H/2$, the proximity force approximation gives
\begin{equation*}
E_{\text{Cas}}^{\text{PFA}}=R_2\int_{-\frac{H}{2}}^{\frac{H}{2}}dz\int_{\pi-\cos^{-1}\frac{R_2}{L}}^{\pi+\cos^{-1}\frac{R_2}{L}} d\phi \,\mathcal{E}_{\text{Cas}}^{\parallel}\left(h(\phi,z)\right),
\end{equation*}
where $L=R_1+R_2+d$, $d$ is the distance between the sphere and the cylinder,   $\mathcal{E}_{\text{Cas}}^{\parallel}\left(d\right)$ is the Casimir energy density between two parallel plates, which is given by
\begin{equation*}
\mathcal{E}_{\text{Cas}}^{\parallel}\left(d\right)=-\frac{\pi^2\hbar c}{1440d^3}
\end{equation*}for Dirichlet or Neumann boundary conditions, and
\begin{equation*}
h(\phi,z)=\sqrt{(R_2\cos\phi+L)^2+(R_2\sin\phi)^2+z^2}-R_1=\sqrt{R_2^2+L^2+2R_2L\cos\phi+z^2}-R_1
\end{equation*}is the distance from the point $(x,y,z)$ on the cylinder to the sphere. After some computations,
we find that for Dirichlet or Neumann boundary conditions, the proximity force approximation gives
\begin{equation}\label{eq12_7_2}
\begin{split}
E_{\text{Cas}}^{\text{PFA}}\sim &-\frac{\pi^3\hbar c}{1440 d^2} R_1\sqrt{\frac{R_2}{R_1+R_2}}.
\end{split}
\end{equation}For perfectly conducting boundary conditions, the leading term is twice of \eqref{eq12_7_2} due to the two polarizations of photons.

By letting the radius of the cylinder approaches infinity, i.e., $R_2\rightarrow \infty$, we obtain the sphere-plane configuration. Indeed, in the limit $R_2\rightarrow\infty$, \eqref{eq12_7_2} gives
$$-\frac{\pi^3\hbar c}{1440 d^2} R_1, $$which is the small separation leading term of the Casimir interaction energy between a sphere and a plane. It is also interesting to compare \eqref{eq12_7_2} to the leading term of the Casimir interaction energy between two spheres of radii $R_1$ and $R_2$, which is
$$-\frac{\pi^3\hbar c}{1440 d^2} \frac{R_1R_2}{R_1+R_2}.$$We see that for all three configurations (sphere-sphere, sphere-cylinder, sphere-plane), the Casimir interaction energy behaves like $\sim 1/d^2$ when $d\ll 1$.

For the next to leading order term, it has been a subject of much interest in these recent few years. For the cylinder-plate configuration, the small separation next to leading order term was computed in \cite{25} from the exact representation of the Casimir interaction energy using perturbation method with careful order counting. This method was then extended to the sphere-plate configuration \cite{44,45,46,47}, the cylinder-cylinder configuration \cite{48} and the sphere-sphere configuration \cite{49}. There is no doubt that this method can be extended to the sphere-cylinder configuration considered here, although some tedious work is required. Less than two years ago, another approach was proposed to compute the small separation next-to-leading order term of the Casimir interaction energy. In \cite{50}, Fosco et al performed derivative expansion on the path integral representation of the Casimir interaction energy and obtained an expression for the next to leading order term in terms of the height profile. They have extended their method in \cite{51, 52} but so far their results can only be applied when one of the objects is planar. Inspired by \cite{50}, Bimonte et al \cite{43} proposed that the Casimir interaction energy has a derivative expansion of the form
 \begin{align*}
 E_{\text{Cas}}^{\text{DE}}=&\int_{\Sigma}d^2\boldsymbol{x} \mathcal{E}_{\text{Cas}}^{\parallel}(H)\Bigl(1+\beta_1(H)\nabla H_1\cdot\nabla H_1+\beta_2(H)\nabla H_2\cdot\nabla H_2\\&\hspace{3cm}+\beta_{\times}(H)\nabla H_1\cdot \nabla H_2+\beta_-(H)\hat{\boldsymbol{z}}\cdot (\nabla H_1\times \nabla H_2)+\ldots\Bigr),
 \end{align*}where $\Sigma$ can be taken to be the $z=0$ plane parametrized by $\boldsymbol{x}=(x,y)$,   $z=H_1(\boldsymbol{x})$ and $z=H_2(\boldsymbol{x})$ are the height profiles of the two objects with respect to $\Sigma$, and $H=H_1-H_2$ is the height difference. The leading term of
 $$\int_{\Sigma}d^2\boldsymbol{x} \mathcal{E}_{\text{Cas}}^{\parallel}(H)$$ is precisely the proximity force approximation.
 Using the invariance of the Casimir interaction energy with respect to tilting the reference plane $\Sigma$, it was found that
 \begin{equation*}
 \begin{split}
 \beta_-(H)=&0,\\
 \beta_{\times}(H)=&\frac{1}{2}\left(1-H\frac{d\log \mathcal{E}_{\text{Cas}}^{\parallel}(H)}{dH}\right)-\beta_1(H)-\beta_2(H)\\
 =&2-\beta_1(H)-\beta_2(H).
 \end{split}
 \end{equation*}
For Dirichlet, Neumann and perfectly conducting boundary conditions,   $\beta=\beta_1=\beta_2$ is found to be a pure number  that only depends on the boundary conditions, which is given by
\begin{equation*}
\begin{split}
\beta^{\text{D}}=&\frac{2}{3},\\
\beta^{\text{N}}=&\frac{2}{3}\left(1-\frac{30}{\pi^2}\right),\\
\beta^{\text{P}}=&\frac{2}{3}\left(1-\frac{15}{\pi^2}\right).
\end{split}
\end{equation*}
The latter implies that up to the next-to-leading order term, the Casimir interaction energy of two perfectly conducting objects is equal to the sum of the Dirichlet and the Neumann Casimir interaction energies in the same geometry \cite{43}.

In the sphere-cylinder configuration that we consider here, we can let $\Sigma$ to be the plane $z=0$, let the sphere  be $x^2+y^2 +(z-L_1)^2=R_1^2$ and the cylinder be $x^2+ (z+L_2)^2=R_2^2$. For simplicity, we assume that $L_1>R_1>0$, $L_2>R_2>0$, and let $L=L_1+L_2$. Notice that $L=R_1+R_2+d$, where $d$ is the distance between the centers of the sphere and the cylinder. Then
 \begin{align*}
 H_1=&L_1-\sqrt{R_1^2-x^2-y^2 },\quad H_2=-L_2+\sqrt{R_2^2-x^2 },\\
 H=& L-\sqrt{R_1^2-x^2-y^2}-\sqrt{R_2^2-x^2},\\
 \nabla H_1\cdot \nabla H_1=&\frac{x^2+y^2}{R_1^2-x^2-y^2},\quad \nabla H_2\cdot \nabla H_2=\frac{x^2}{R_2^2-x^2},\\
 \nabla H_1\cdot \nabla H_2=&-\frac{x^2}{\sqrt{R_1^2-x^2-y^2}\sqrt{R_2^2-x^2}}.
 \end{align*}
After some computations, we find that
 \begin{equation*}
 \int_{\Sigma}d^2\mathbf{x}\frac{1}{H^3}=\frac{\pi R_1}{d^2}\sqrt{\frac{R_2}{R_1+R_2}}\left(1-\frac{d}{R_1R_2(R_1+R_2)}\left(\frac{3}{8}R_1^2 +R_2^2\right)+\ldots\right),\end{equation*}\begin{equation*} \begin{split}
 \int_{\Sigma}d^2\mathbf{x}\frac{1}{H^3}\nabla H_1\cdot\nabla H_1=&  \frac{\pi}{d}\sqrt{\frac{R_2}{(R_1+R_2)^3}}(R_1+2R_2)+\ldots,\\
  \int_{\Sigma}d^2\mathbf{x}\frac{1}{H^3}\nabla H_2\cdot\nabla H_2 =&  \frac{\pi}{d}\sqrt{\frac{1}{R_2(R_1+R_2)^3}} R_1^2+\ldots,\\
  \int_{\Sigma}d^2\mathbf{x}\frac{1}{H^3}\nabla H_1\cdot\nabla H_2 =& \frac{\pi}{d}\sqrt{\frac{R_2}{(R_1+R_2)^3}} R_1 +\ldots.
 \end{split}\end{equation*}
 Therefore, derivative expansion shows that up to the next to leading order term,
 \begin{equation*}\begin{split}
 E_{\text{Cas}}^{\text{DE}}=&-\frac{\alpha \pi^3\hbar c R_1}{1440 d^{2}}  \sqrt{\frac{R_2}{R_1+R_2}}\left(1 - \frac{d}{R_1R_2(R_1+R_2)}\left(\frac{3}{8}R_1^2+R_2^2\right)
 \right.\\&\hspace{2cm}\left.+\beta d\left(\frac{R_1+2R_2}{R_1(R_1+R_2)}+\frac{R_1 }{R_2(R_1+R_2)}\right)-(2-2\beta)d\frac{1}{R_1+R_2} \right)\\
 =&-\frac{ \alpha\pi^3\hbar c R_1}{1440 d^{2}}  \sqrt{\frac{R_2}{R_1+R_2}}\left(1 - \frac{5}{8}\frac{d}{R_1+R_2}+\left(2\beta-1\right)\frac{d}{R_1}+\left(\beta-\frac{3}{8}\right)\frac{d}{R_2} \right),
 \end{split}\end{equation*} where $\alpha^{\text{D}}=\alpha^{\text{N}}=1$ and $\alpha^{\text{P}}=2$.
More precisely, we have
\begin{equation*}
\begin{split}
E_{\text{Cas}}^{\text{DE}, \text{D}}=&-\frac{  \pi^3\hbar c R_1}{1440 d^{2}}  \sqrt{\frac{R_2}{R_1+R_2}}\left(1 - \frac{5}{8}\frac{d}{R_1+R_2}+ \frac{d}{3R_1}+ \frac{7d}{24R_2} \right),\\
E_{\text{Cas}}^{\text{DE}, \text{N}}=&-\frac{  \pi^3 \hbar c R_1}{1440 d^{2}}  \sqrt{\frac{R_2}{R_1+R_2}}\left(1 - \frac{5}{8}\frac{d}{R_1+R_2}+\left(\frac{1}{3}-\frac{40}{\pi^2}\right) \frac{d}{R_1}+ \left(\frac{7}{24}-\frac{20}{\pi^2}\right)\frac{ d}{ R_2} \right),\\
E_{\text{Cas}}^{\text{DE}, \text{P}}=&-\frac{  \pi^3 \hbar cR_1}{720 d^{2}}  \sqrt{\frac{R_2}{R_1+R_2}}\left(1 - \frac{5}{8}\frac{d}{R_1+R_2}+\left(\frac{1}{3}-\frac{20}{\pi^2}\right) \frac{d}{R_1}+ \left(\frac{7}{24}-\frac{10}{\pi^2}\right)\frac{ d}{ R_2} \right).
\end{split}
\end{equation*}
It will be interesting to compare these results to the results computed directly from the exact formulas obtained in Section \ref{sec2} and Section \ref{sec3}. This will be considered in a future work.

\end{document}